\documentclass[
 reprint,
 amsmath,amssymb,
 aps,onecolumn,
]{revtex4-2}

\usepackage{graphicx}
\usepackage{dcolumn}
\usepackage{bm}
\usepackage{amsmath}
\usepackage{amssymb}
\usepackage{soul}
\usepackage{ulem}
\usepackage{siunitx} 
\usepackage{upgreek}
\usepackage[mathlines]{lineno}
\usepackage{MnSymbol}
\usepackage{hyperref}

\usepackage{tikz}

\begin{document}

\title{Impact of the formation angle on the drag of bio-inspired $\pmb \vee$-formations}

\author{Prasoon Suchandra}
\affiliation{Department of Mechanical Engineering, Georgia Institute of Technology, Atlanta, GA 30332}

\author{Shabnam Raayai-Ardakani}
 \email{shraayai@bu.edu}
\affiliation{Department of Mechanical Engineering, Boston University, Boston, MA 02215
}

\begin{abstract}
Bio-inspired $\pmb \vee$ flight formation is a well known technique for energy saving among groups of fixed-wing aircraft, and as of recently, for groups of quad-rotors. Here, we study the effect of the formation angle on the performance of each of the members of a 5-member $\pmb \vee$-formation in terms of the flow field, and drag force. We employ axisymmetric cylinders, which are non-lifting in solo condition to reduce/eliminate the effect of the lift (lateral force) on the group performance, and use time-resolved, multi-illumination, consecutive-overlapping particle image velocimetry (PIV) to capture the velocity field around and in-between the members. Over a range of $\pmb \vee$-formation angles, we see various degree of drag reduction, with the highest drag reduction ($\sim 80\%$) for the interior members of the tightest formation (formation with the smallest $\pmb \vee$-angle and the most overlap in frontal views). All formation members experience some levels of drag reduction up for $\pmb \vee$-angle of around $50^{\circ}$ and in formation with $\pmb \vee$-angle greater than $50^{\circ}$, only the leading member experiences observable drag reduction. We explore the complex flow dynamics between the formation members in terms of wake-body and wake-wake interactions, and the bleeding (gap) flow. We present the mean and fluctuating quantities, as well as the dynamics of the vortex shedding and circulation in the wakes of the members, and discuss how these flow characteristics relate to the drag of each member, both as a function of their position within the $\pmb \vee$ and the angle of the formation. This current study serves as a baseline for further explorations of wake-body and wake-wake interactions of flow past groups of bodies, and demonstrates how changing formation angle can help achieve a desired group performance (like minimum drag).
\end{abstract}

                             
\maketitle

\section{Introduction}\label{sec:intro}

$\pmb \vee$-formation is a common travel pattern seen among groups of migratory birds, including but not restricted to Canada geese \cite{Gould_1974, may1979flight, cutts1994energy}, pelicans \cite{Weimerskirch_2001}, pigeon flocks \cite{nagy2010hierarchical}, and Northern bald ibises \cite{portugal2014upwash} with recorded energetic and psychological benefits for the birds \cite{Weimerskirch_2001}. Formation flight of fixed wing aircraft, especially $\pmb \vee$-formations, have also shown energetic benefits compared with a single aircraft, with more military implementation of this flight strategy than commercial ones. More generally, group motion is a common behavior among animals, examples ranging from insect swarms \citep{attanasi2014collective,cavagna2017dynamic,mendez2018density}, schools of fish \citep{liao2003fish,Zhang_2023,Ko_2023,Ito_2023,kelly2024} to other aquatic life-forms \citep{bill1976drag}. Geometric arrangements in flow conditions are also observed in mangrove vegetation patches in riverfront and coastal areas which are able to control flood and prevent soil erosion \citep{Kazemi_2018, Gymnopoulos_2019,Kazemi_2021,Liu_2021,Ferreira_2021,He_2024}. Studies of flow past solid arrays are also essential for engineering applications, such as heat exchangers in power plants \citep{Sayers_1987,Stanescu_1996,Kumar_2021}, and designs of marine structures \citep{Sayers_1987,Yagci_2021,He_2024}.  Different speed sports such as speed skating and cycling use different drafting techniques in competitions as well \cite{van2023drafting, Blocken_2018}.

Formation flights of birds and fixed wing aircraft has been mainly explored based on the potential flow analysis  of the wing-tip vortices of a finite-span lift-generating body and the lift-induced upwash outside the wake \cite{lissaman1970formation, hummel_1983, hummel1996use, blake1998design, jacques2001analytical, kawabe_2007, xu2014aircraft, ning2011aerodynamic, bower2009formation}. However, in recent years, application of formation strategies for drone swarms has shown promise in managing and controlling the battery consumption of groups of rotorcraft \cite{guo2023drone}. With the differences in the flight mechanism of the fixed and rotary-wing systems, the potential flow model of formation based on lift-induced upwash does not readily apply to the new generation of vehicles, including those in air and in water. Previously, we have shown that using non-lift-generating cylinders in a $\pmb \vee$-formation allows us to alter the drag force experienced by the members of a formation. Using two formation angles of 37.6$^{\circ}$ (narrow) and $67.6^{\circ}$ (wide), we have reported the narrow formation, where members have frontal overlaps, is able to reduce the drag force experienced by all the bodies, while with the wide formation, we were only able to reduce the drag of the leading member of the formation \cite{suchandra2024}. Therefore, understanding the mechanics of flow around arrangements of objects, specifically with a focus on the impact of the variations in the geometry of the formation on the dynamics response of the flow, can be effectively used as an optimization tool for tuning the aero/hydro-dynamic responses, like minimizing energy expenditure for a group of vehicles. 

Here, we study the effect of the $\pmb \vee$-angle on the aero-/hydro-dynamic performance of $\pmb \vee$-formation of 5 non-lifting members, i.e. cylinders. We explore the dynamics of the group in terms of the wake-wake and wake-body interactions between the different members and how this impacts the drag force each member is experiencing. We use cylinders as canonical models of non-lifting objects \citep{Williamson_1996}. We focus on 5-member groups as an intermediate representatives that captures many of the qualities of 3-member and 7-member dynamics as previously discussed \cite{suchandra2024}. While some of the flow patterns in formations with 5 and 7 members \cite{suchandra2024} show similarities with the patterns previously studied for side-by-side cylinders \cite{williamson1985evolution, sumner2000flow, sumner2010two}, as well as cylinders in tandem, and staggered arrangements of 2 or 3 cylinders \cite{zdravkovich1977interference,zdravkovich1977review,zdravkovich1987effects,sumner2010two,zhou2016wake,rockwell1998vortex} the qualitative interference patterns are not sufficient to describe the flow interactions in formations with a larger number of members \cite{zdravkovich1987effects}, including $\pmb{\vee}$-formations. We present a high resolution and quantitative experimental exploration and employ a multi-light-sheet \cite{fu2023multisheet, suchandra2024} approach to overcome the challenge of shadows in the presence of multiple bodies which has limited the previous studies to views of flows on only one side, or the wakes \cite{Kazemi_2018,Nair_2023}. We also recognize the challenges associated with implementation and use of index-matched facilities, including limited material availability, high costs, and the potential hazards of specialized fluids which make such facilities impractical \cite{amini2012investigation, wright2017review} and use the multi-light-sheet approach as a cost-effective alternative to index-matching. In addition, we use a consecutive-overlapping imaging approach which allows us to capture a larger area of interest around the formations at high resolution \cite{suchandra2024,Fu_2023,fu2023multisheet, fu2025localised}. 

This paper is organized as follows: in section \ref{sec:methods}, we give an overview of the experimental setup and procedures, and discuss the results in section \ref{sec:results}. In section \ref{sec:drag}, we present the drag coefficients of each member as a function of their location within the array and the $\pmb \vee$-formation angle $\phi$, and discuss the general trends. In sections \ref{sec:global} and \ref{sec:local}, we continue to explore the velocity and pressure distribution around the members and how the variations in the flow fields affect the forces experienced by each member as the formation angle is changed and how does this affect the overall performance of the group. In section \ref{sec:vortexshedding}, we present a preliminary overview of some of our observations in the potential changes in lift experienced by the bodies, and the temporal variation of the circulation in the wakes and the vortex shedding pattern. Eventually in section \ref{sec:conclusion}, we highlight the main takeaways from the current study.

\section{Methods}\label{sec:methods}

\subsection{Geometric setup}\label{sec:geometry}

\begin{figure}[htpb]
    \centering
    \includegraphics[width=0.95\textwidth]{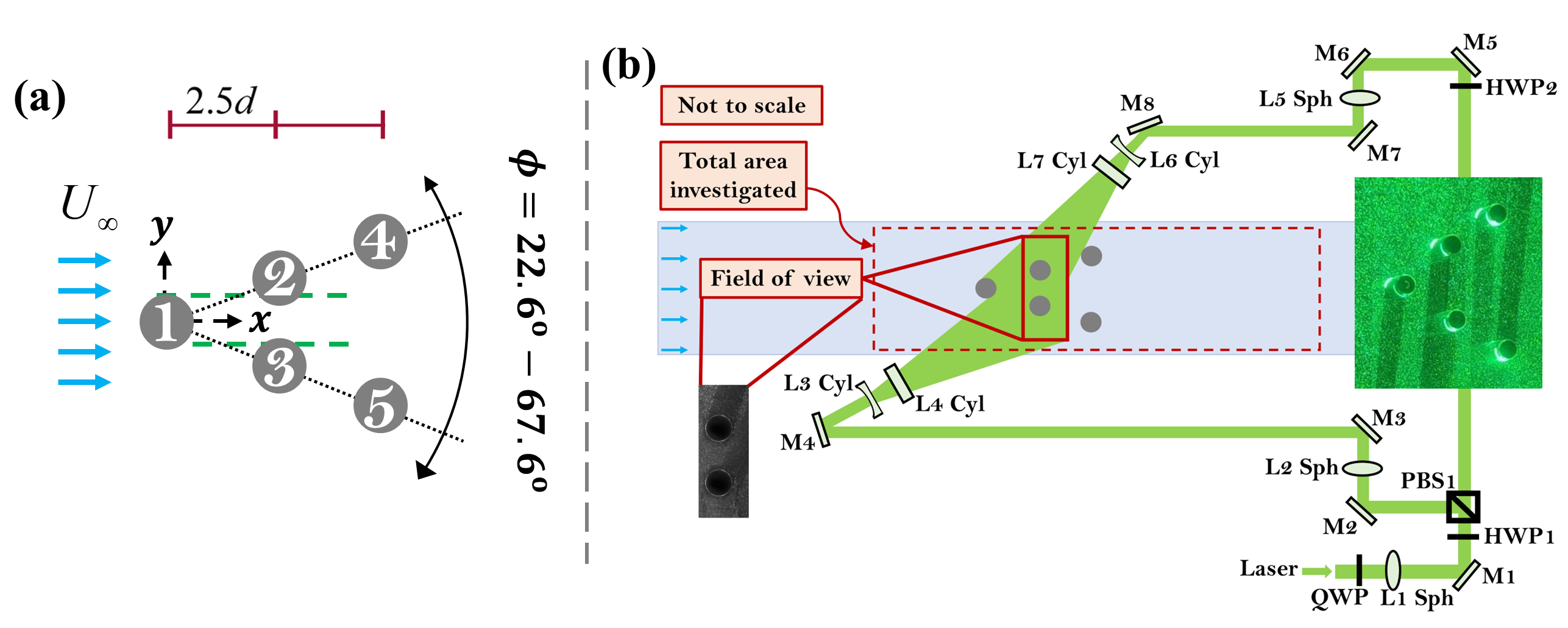}
    \caption{(a) A schematic summary of the $\pmb \vee$-formations, along with the member numbering and the coordinate system. (b) Schematic of an angled double-sheet laser illumination for visualizing the flow inside and around a multi-body sample. The view of the intersection of the two light sheets around the array members is also shown from the bottom of the test section.}\label{fig:cases}
\end{figure}

To study the dynamics of flow around formations of non-lifting objects, the $\pmb \vee$-formations here are designed as a symmetric, two-echelon linear configuration, where the formation angle of the $\pmb \vee$, called $\phi$, is the angle between the two echelons or branches \cite{suchandra2024}. Here, we consider a group of 5 stationary non-lifting circular cylinders of diameter $d = 6$ mm, arranged in $\pmb \vee$-formations (Fig.~\ref{fig:cases}(a)). We keep the horizontal distance between the rows of the members constant at $2.5d$ which is slightly lower than the vortex formation length \citep{Williamson_1996, bearman1965investigation, bearman1969vortex, bearman1984vortex, Unal_1988,Chopra_2019,suchandra2024} of a single cylinder ($\sim 2.6d$) at the given conditions and within the region of ``wake interference'' \cite{zdravkovich1987effects}. Due to the physical limit, the lowest possible $\phi$ is $22.6^{\circ}$, where members 2 and 3 are in contact which corresponds to the ``touching line'' as previously noted in studies of 2 and 3 cylinders/pipes in tandem \cite{zdravkovich1987effects}. Independent of the results, we divide the experimental cases here into two categories, with the first one comprised of the formations where the members have streamwise overlaps along each echelon, or $\phi < \phi^{*} = 43.6^{\circ}$, and the second group consisting of formations with no frontal overlaps among the members, corresponding to $\phi > \phi^{*} = 43.6^{\circ}$. This streamwise or frontal extent is indicated by green dashed lines for members 1, 2, and 3 in Fig.~\ref{fig:cases}(a). We extend the investigation, in steps of $5^{\circ}$, till $\phi = 67.6^{\circ}$ where the flow past all the members almost (but not quite) resembles that of 5 solo members \citep{suchandra2024}. Overall, we present 10 cases in total and explore how the change in $\phi$ can help to achieve a specific desired group performance (like minimum drag). 

\subsection{Experimental procedure}\label{sec:exptproc}

Our experiments are conducted in a water tunnel facility \citep{Fu_2023,fu2023multisheet,suchandra2024,Fu_2025}, with a test section of 20 cm $\times$ 27 cm in cross-section and 2 m in length. The water height is kept at about 20 cm during the experiments. All experiments are performed at free-stream speed of $U_{\infty} \approx 17.7$ cm/s with run-to-run variation of $\pm 0.3$ cm/s. The Reynolds number based on this free-stream speed and cylinder diameter $d$ is ${\rm Re}_d \approx 1100$, corresponding to a turbulent wake behind a single circular cylinder \citep{Lienhard_1966,vandyke_1982}. 

The velocity field around the sample is obtained via a time-resolved two-dimensional, two-component (2D-2C) particle image velocimetry (PIV) system \citep{adrian2011particle, Adrian_1991, Raffel_2018}, as shown in Fig.~\ref{fig:cases}(b). It consists of a double-pulsed Nd:YLF laser (Terra PIV, Continuum Laser) of 527 nm (green light), a high-speed camera (Chronos 2.1, Kron Technologies Inc.) at a resolution of 720 $\times$ 1920 pixels with a 100 mm macro lens (Canon EF 100 mm f/2.8L Macro Lens), and a timing unit (Arduino Teensy Board \cite{fu2023multisheet, teensy}) synchronizing the laser pulses and camera capture. The imaging magnification is set to each pixel capturing about 18 $\rm \upmu m$. The delay between the two successive laser pulses for the PIV is set at $\delta t$ = 1000 $\rm \upmu s$. The laser is operated at a frequency of 500 Hz (note that the vortex shedding frequency of a single cylinder at ${\rm Re}_d \approx 1100$ is around 6 Hz). Water is seeded with 8-12 $\rm \upmu m$ hollow glass particles (TSI Inc.) as PIV tracers.  

To illuminate the field of view around the array members, we use an angled double-light sheet strategy as shown in Fig.~\ref{fig:cases}(b), which is a slight variation of the double and quadruple light sheet strategies of \citet*{Fu_2023,fu2023multisheet}. The optical components are arranged such that the incoming laser beam is divided into two beams using a half-wave plate (HWP) and a polarizing beam splitter (PBS) and directed toward the front and back of the tunnel through multiple mirrors (M). Additionally, a HWP and a quarter-wave plate (QWP) are used to get the proper polarity of the beams on the two sides, getting the best possible illumination in the camera's field of view. The beams are passed through a combination of spherical (Sph) and cylindrical (Cyl) lenses, illuminating the field of view as demonstrated in Fig.~\ref{fig:cases}(b) (Note that the combination of laser optics used depends on the type of laser, camera and the intended application). Each laser sheet is about 2 mm in thickness. Formations with small $\phi$ (i.e., those with streamwise overlaps between members; $\phi < 43.6^{\circ}$) result in dark spots between various members (in particular, between members 2 and 3). In such cases, with a slight change in the angle of the light sheet, we can illuminate the dark area while placing a different region in shadow \citep{suchandra2024}. In these cases, the field of view is imaged more than once so that each portion of the field of view is illuminated (without shadow) in at least one set of images \citep{suchandra2024}. 

The camera is traversed throughout the total area of interest via a computer numerically controlled (CNC) carrier in consecutive-overlapping steps \citep{Fu_2023, fu2023multisheet}. The imaging is repeated across the streamwise length ($x$-direction) with about 30\% overlap and the normal width ($y$-direction) with about 20-50\% overlap (case-to-case variations) with the previous position \citep{suchandra2024}. The laser sheet optics are moved manually. PIV is performed for image-pairs captured at each location and the results are stitched to get the full velocity field. At each field of view, 990 PIV image pairs are captured, which are found to be sufficient for the convergence of the mean and fluctuation quantities \citep{suchandra2024}.

The PIV image pairs are processed using the open-source software OpenPIV \citep{OpenPIV}. The PIV search area is 64 px $\times$ 64 px in size and the interrogation window is 32 px $\times$ 32 px with 87.5\% overlap (28 px) with the neighboring windows. The resolution of the processed data is $\sim$ 80 $\rm \upmu m$ per vector. Spurious PIV data are detected and removed using the universal outlier detection method of \citet{Westerweel_2005}.

\subsection{Force calculations}\label{sec:force_cal}

Using the PIV data, we calculate the drag forces experienced by each member using a control volume analysis, as discussed in detail by \citet{suchandra2024}. In summary, we apply the conservation of mass and momentum fluxes through an appropriately chosen control volume (CV) around an entire formation as well as each of its members and use the Reynolds-averaged integral momentum (RAIM) conservation equation \citep{Ferreira_2021,Fu_2023,suchandra2024} as given by 

\begin{equation}\label{eq:RAIM}
\begin{aligned}
D\pmb{e}_x + L\pmb{e}_y \, = \, & \underbrace{- \, \rho \int_{\rm CS} \overline{(\pmb{u} + \pmb{u'}) [(\pmb{u} + \pmb{u'}) \cdot \pmb{n}]} dA \, + \, \pmb{D}_{\rm{3D}}}_\text{Momentum fluxes and Reynolds stresses} \,
& \underbrace{+ \, \int_{\rm CS} \pmb{n} \cdot \overline{\pmb{\tau}} dA}_\text{Viscous forces} \, \underbrace{- \, \int_{\rm CS} p \, \pmb{n} \, dA}_\text{Pressure forces} 
\end{aligned}
\end{equation}

\noindent which provides the drag and lift forces (on the body enclosed in the CV) per unit length of the cylinder, $D$ and $L$, respectively and $\pmb{e}_x$ and $\pmb{e}_y$ are the unit vectors in each of the corresponding directions, and $\rm CS$ denotes the control surface. Here, $\rho$ and $\mu$ are the density and dynamic viscosity of the fluid (water) respectively, $\overline{(...)}$ denotes ensemble-averaging, $\pmb{\tau}$ is the stress tensor tensor, $p$ is the pressure, and $\pmb{n}$ is the local outward unit normal vector to the control surface. We employ the Reynolds decomposition \citep{Kundu_2008} and decompose the instantaneous velocity vector $\mathbf{\tilde{u}}(x,y,t)$ into the mean $\mathbf{u}(x,y)$ and the fluctuating component $\mathbf{u'}(x,y,t)$, as $\mathbf{\tilde{u}}(x,y,t) = \mathbf{u}(x,y) + \mathbf{u'}(x,y,t)$. In the current work, contribution of the shear stresses to the viscous forces on the outer boundaries of the CV are not neglected as these boundaries are sufficiently close to the enclosed solid as previously discussed \cite{suchandra2024}, especially when calculating the drag force on an individual member of the formation. As for the pressure distribution, we use the velocity field from the PIV and line integration \citep{Liu_2006, Charonko_2010, Kat_2013, Oudh_2013, Liu_2016, Liu_2020, Nie_2022} of the pressure gradient terms in the Reynolds-averaged Navier-Stokes (RANS) equations \citep{Lumley_1972,Pope_2000,Kundu_2008} to obtain the mean pressure as given by

\begin{equation}\label{eq:px}
\begin{split}
p(x) - p(x_{\rm ref}) = \int_{x_{\rm ref}}^{x} \biggl[ -\rho \left(u \frac{\partial u}{\partial x} + v \frac{\partial u}{\partial y} \right) + \mu \left( \frac{\partial^2 u}{\partial x^2} + \frac{\partial^2 u}{\partial y^2} \right) - \rho \left(\frac{\partial \overline{u' u'}}{\partial x} + \frac{\partial \overline{u' v'}}{\partial y} \right) \biggl] dx 
\end{split}
\end{equation}

\begin{equation}\label{eq:py}
\begin{split}
p(y) - p(y_{\rm ref}) = \int_{y_{\rm ref}}^{y} \biggl[ -\rho \left(u \frac{\partial v}{\partial x} + v \frac{\partial v}{\partial y} \right)
+ \mu \left( \frac{\partial^2 v}{\partial x^2} + \frac{\partial^2 v}{\partial y^2} \right) - \rho \left(\frac{\partial \overline{u' v'}}{\partial x} + \frac{\partial \overline{v' v'}}{\partial y} \right) \biggl] dy.
\end{split}
\end{equation}

\noindent where $u$ and $v$ are the streamwise and normal components of the velocity vector respectively. Using the above analysis, we employ at least 20 different control volumes around each cylinder, and report the mean values and their 95\% confidence intervals in the upcoming figures.

\subsection{Bleeding flow}\label{sec:bleed}

The flow passing through the spacing between formation members, called the ``bleeding flow'' or the ``gap flow'' \cite{Chang_2015, Zhou_2019, Nicolai_2020, Liu_2021, Yagci_2021, sumner2000flow, sumner2010two}, has a significant effect on the forces experienced by the members \citep{suchandra2024}. Here, we mark the mean velocity of the flow through the gap between the members as $\boldsymbol{u}^b_{ij}$ with $i$ and $j$ referring to the relevant member numbers. We also define the equivalent bleeding flow, $U_{ij}^b$, between any two array members $i$ and $j$ which is calculated by dividing the total volumetric flow rate (per unit length of the cylinder) between the two members by the linear distance between those two members, as given by

\begin{equation} 
U_{ij}^b = \frac{ \left \vert \int_{i}^{j} \pmb{u}_{ij}^b \cdot \pmb{n} \ dA \right \vert}{\int_{i}^{j} dA} = \frac{\left \vert\int_{i}^{j} \pmb{u}_{ij}^b \cdot \pmb{n} \ L_z \ d\ell_{ij}\right \vert}{\int_{i}^{j} L_z \ d\ell_{ij}} = \frac{\left \vert \int_{i}^{j} \pmb{u}_{ij}^b \cdot \pmb{n} \ d\ell_{ij}\right \vert}{\ell_{ij}},
\end{equation}

\noindent where $dA = L_z \ d\ell_{ij}$ is the infinitesimal area element ($L_z$ is the length of a member in the $z$ direction, perpendicular to $xy$-plane), $\pmb{n}$ is the normal to the area element, $\left \vert... \right \vert$ denotes the magnitude, and $\ell_{ij}$ is the shortest linear distance between the surfaces of the two members.

\section{Results}\label{sec:results}

\subsection{Drag force experienced by each member}\label{sec:drag}

\begin{figure}[!b]
    \centering
    \includegraphics[width=1\textwidth]{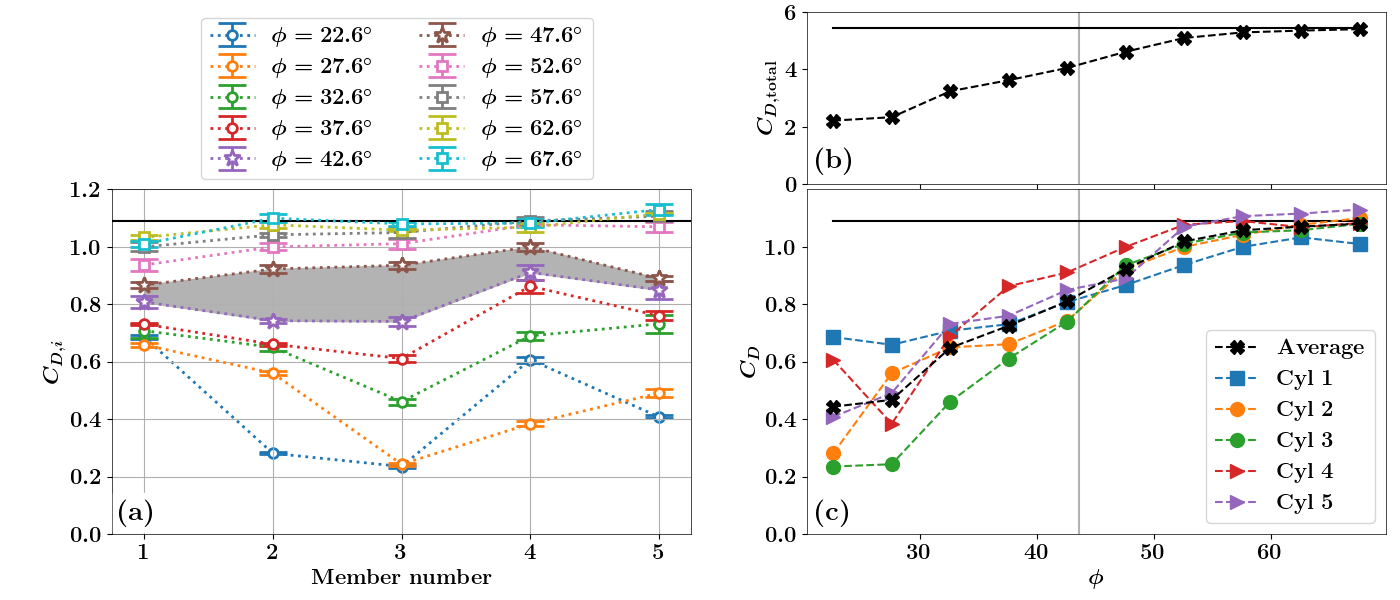}
    \caption{(a) Ensemble-averaged drag coefficient, $C_{D,i}$, for all the individual members of all the $\pmb \vee$-formations as a function of the member number. Error bars are based on 95\% confidence intervals. Overlapping formations are designated with round symbols, non-overlapping formations are shown with square symbols, and the two formations close to the cross-over $\phi^*=43.6^{\circ}$ are marked with star symbols and the separation is colored in light gray for visibility. (b) Total ensemble-averaged drag coefficient, $C_{D,\rm{total}}$, against angle of $\pmb \vee$, $\phi$, for all the formations. The cross-over angle $\phi^*=43.6^{\circ}$ is marked with gray vertical line. (c) Drag coefficient, $C_{D,i}$, of part (a) re-plotted as a function of the formation angle $\phi$. Members of the same row are shown with the same symbols and different colors. The black dashed line corresponds to the average drag of the group, $C_{D,\rm{total}}/5$.}\label{fig:CD_t_i}
\end{figure}

We start by looking at the drag forces experienced by each of the individual members of the different $\pmb \vee$-formations, as indicated by $C_{D,i}= D_i/(0.5 \rho U_{\infty}^2 d)$ for member $i$, and the drag of the entire formation as one group, denoted by $C_{D, {\rm total}}$, presented in Fig.~\ref{fig:CD_t_i}. We present the drag force of individual members as a function of their member numbers (i.e. positions) in Fig.~\ref{fig:CD_t_i}(a) for the various formations, the total drag of the entire formation as a function of the formation angle in Fig.~\ref{fig:CD_t_i}(b), and also re-plot the drag of individual members, this time as a function of the formation angle in Fig.~\ref{fig:CD_t_i}(c) for clarity. For easier comparison across the cases, the denominator of the drag coefficients are kept as a constant according to the \textit{a priori} knowledge of the geometry and the flow $(0.5 \rho U_{\infty}^2 d)$. The trends in the drag force are seen both as a function of the angle of the formation, and the position of the member presented with numbers shown on the cylinders in the Fig.~\ref{fig:cases}(a) and upcoming figures. At a first glance, as seen in Fig.~\ref{fig:CD_t_i}(a), the two categories of overlapping and non-overlapping formations defined in Sec.~\ref{sec:geometry}, separate the drag responses into two groups where on the figure, the space between the two cases that are at the margin of the overlap, i.e. $\phi=42.6^{\circ}$ and $\phi=47.6^{\circ}$ is colored in gray. But after a careful observation, the pattern of the drag force of individual members of $\phi = 47.6^{\circ}$ formation shows a more similar trend to the overlapping cases than the non-overlapping cases. 

In formations with $\phi\leqslant 47.6^{\circ}$, all members experience substantial reductions in the drag force, with asymmetries seen in the $C_D$ of members of the same row of the same formation (comparing 2 against 3, and 4 against 5). At lower angles with larger overlaps between the frontal views of the members, especially $22.6^{\circ} \leqslant \phi \leqslant 37.6^{\circ}$, the flow tends to take asymmetric patterns (asymmetry with respect to the $y=0$ line of the formation) resulting in members of the upper and lower echelons experiencing different drag forces and will be discussed further in Secs.~\ref{sec:global} and \ref{sec:local}. For formation with $\phi\geqslant52.6^{\circ}$, we see less or no reductions in the drag force compared with the case of the single cylinder, where the main benefit (in terms of drag reduction) goes to the leading member 1 of the formation. For these cases, the trend in the drag coefficient is more uniform, with the lowest drag going to member 1, then slightly larger drag measured for members 2 and 3, and then the largest values going to members 4 and 5 (see Fig.~\ref{fig:CD_t_i}(a)). 

The total drag force experienced by the entire formation (Fig.~\ref{fig:CD_t_i}(b)) increases monotonically with the formation angle. For the formations with $\phi \lessapprox 60^{\circ}$, the total drag is less than the total drag if each member behaved like an isolated cylinder, i.e., $C_{D,\textnormal{total}} < 5 \times C_D^{\textnormal{iso cyl}}$. Thus, even without streamwise overlaps, $\pmb \vee$-formations can offer a lower total drag and such non-overlapping formations are viable options for managing the energy consumption within groups. The increase in the angle of the formation, results in a consistent increase in the drag force each member experiences, except for one point on the figure which does not follow this trend (member 4 in $\phi=22.6^{\circ}$ and $27.6^{\circ}$ formations), mainly due to the asymmetries seen in $\phi=22.6^{\circ}$ and $\phi=27.6^{\circ}$ cases. For the leading member 1, the drag force takes a mild increasing trend, starting from $C_{D,1} \approx 0.67$ for $\phi \leqslant 27.6^{\circ}$, then increasing monotonically for $ 32.6^{\circ} \leqslant\phi \leqslant 52.6^{\circ}$, and asymptotically reaching a $C_{D,1} \approx 1$ for $\phi \geqslant 57.6^{\circ}$. In addition, the leading member takes advantage of the presence of members 2 and 3 in the second row, which are directly in the path of the wake of member 1. This results in some levels of pressure recovery downstream (or equivalently, the leading member receiving a ``forward push" from downstream members 2 and 3)\citep{suchandra2024}, leading to a reduced drag compared to a solo cylinder case (more in Secs.~\ref{sec:global} and \ref{sec:local}). 

As seen in Fig.~\ref{fig:CD_t_i}(c), member 1 experiences the largest drag of the group at lower angles, and crosses over to less than the trailing members and more than the $C_D$ of the middle row at intermediate angles (all overlapping), and for all non-overlapping angles it has the lowest drag of the formation. As the angle of the formation increases, especially past $\phi^*$, we see the drag force of interior and trailing members asymptotically get closer to the drag of a solo case. This is especially visible for members 2 and 3 in $\phi>60^{\circ}$, and members 4 and 5 in $\phi>50^{\circ}$, with member 5 recording a slightly higher drag force than the solo case at the two largest formation angles tested. Member 1, on the other hand, maintains a lower drag force than the solo case. The lowest drag force among all member and cases tested was recorded for member 3 of the $\phi = 22.6^{\circ}$ and $27.6^{\circ}$ formations ($C_D \approx 0.24$) with member 2 of the $\phi=22.6^{\circ}$ coming in as a close second ($C_D \approx 0.28$).

\subsection{Flow pattern through $\pmb{\vee}$-formations}\label{sec:global}

\begin{figure}[!ht]
    \centering
    \includegraphics[width=1\textwidth]{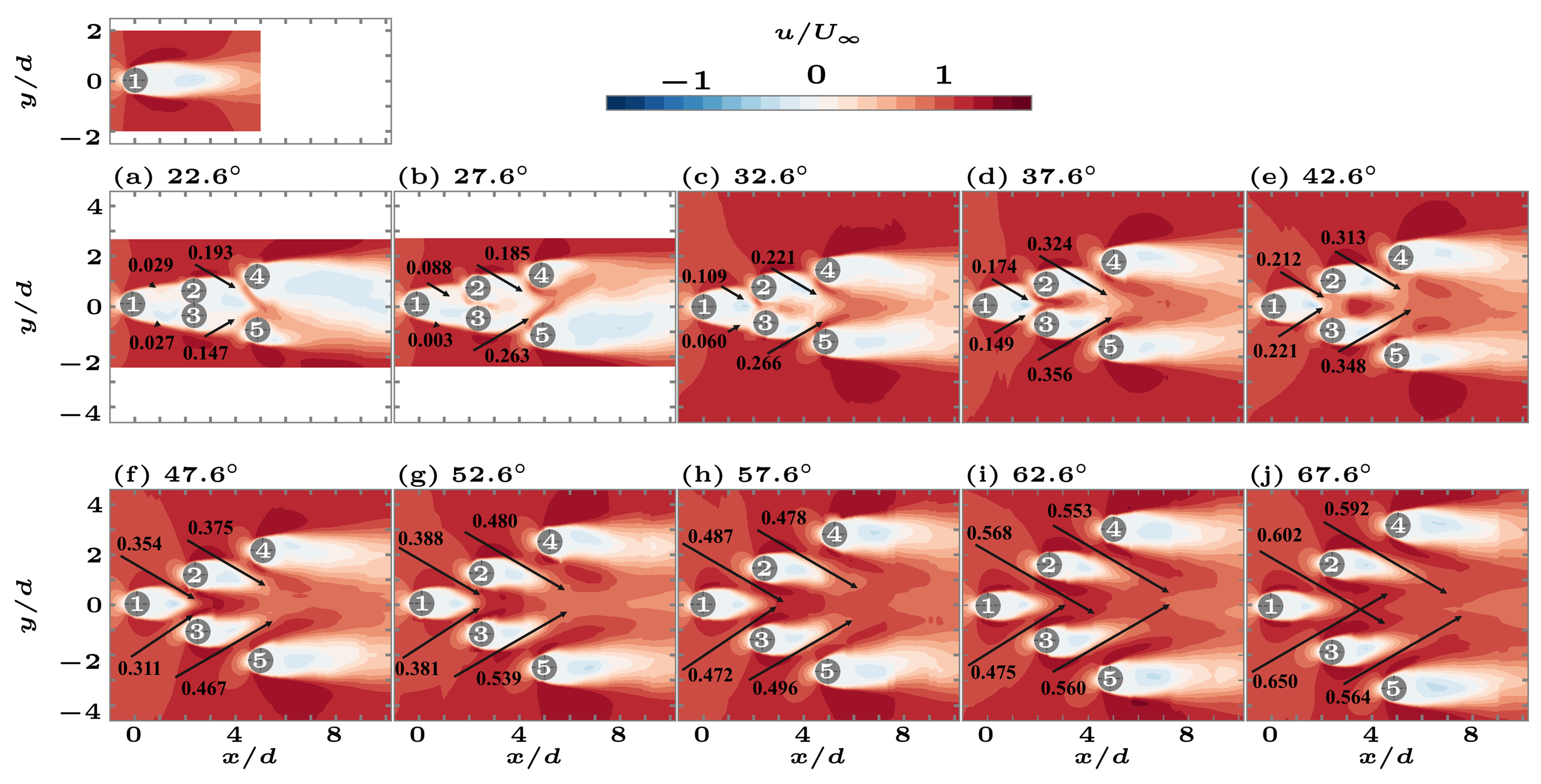}
    \caption{Contours of normalized mean streamwise velocity, $u(x,y)/U_{\infty}$ for formations with $\phi$ going from $22.6^{\circ}$ to $67.6^{\circ}$ in steps of $5^{\circ}$. Sub-figures (a-e) correspond to the cases with frontal overlaps and sub-figures (f-j) correspond to the cases with no frontal overlaps. The velocity contours for the solo cylinder is presented on the top as reference. The magnitude of the bleeding flow, $U_{ij}^b/U_{\infty}$, between the members of each echelon of the formation are shown on the figures. Arrows are shown to indicate inward and outward flow and are not accurate representation of the average direction of the flow. Arrows' lengths are proportional to the magnitude of the bleeding flow.}\label{fig:U}
\end{figure}

\begin{figure}[!ht]
    \centering
    \includegraphics[width=1\textwidth]{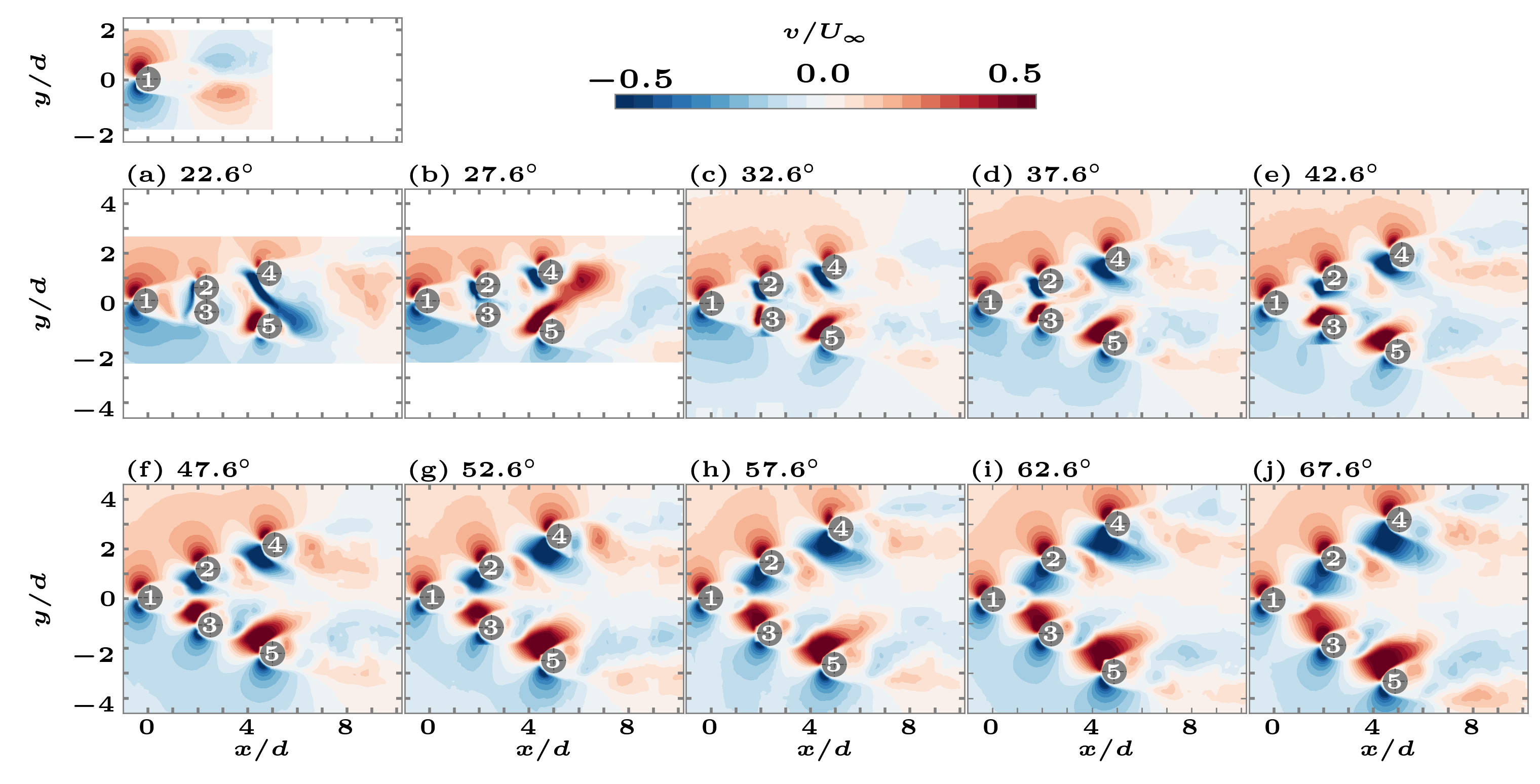}
    \caption{Contours of normalized mean normal velocity, $v(x,y)/U_{\infty}$ for formations with $\phi$ going from $22.6^{\circ}$ to $67.6^{\circ}$ in steps of $5^{\circ}$. Sub-figures (a-e) correspond to the cases with frontal overlaps and sub-figures (f-j) correspond to the cases with no frontal overlaps. The velocity contours for the solo cylinder is presented on the top as reference.}\label{fig:V}
\end{figure}

\begin{figure}[!ht]
    \centering
    \includegraphics[width=1\linewidth]{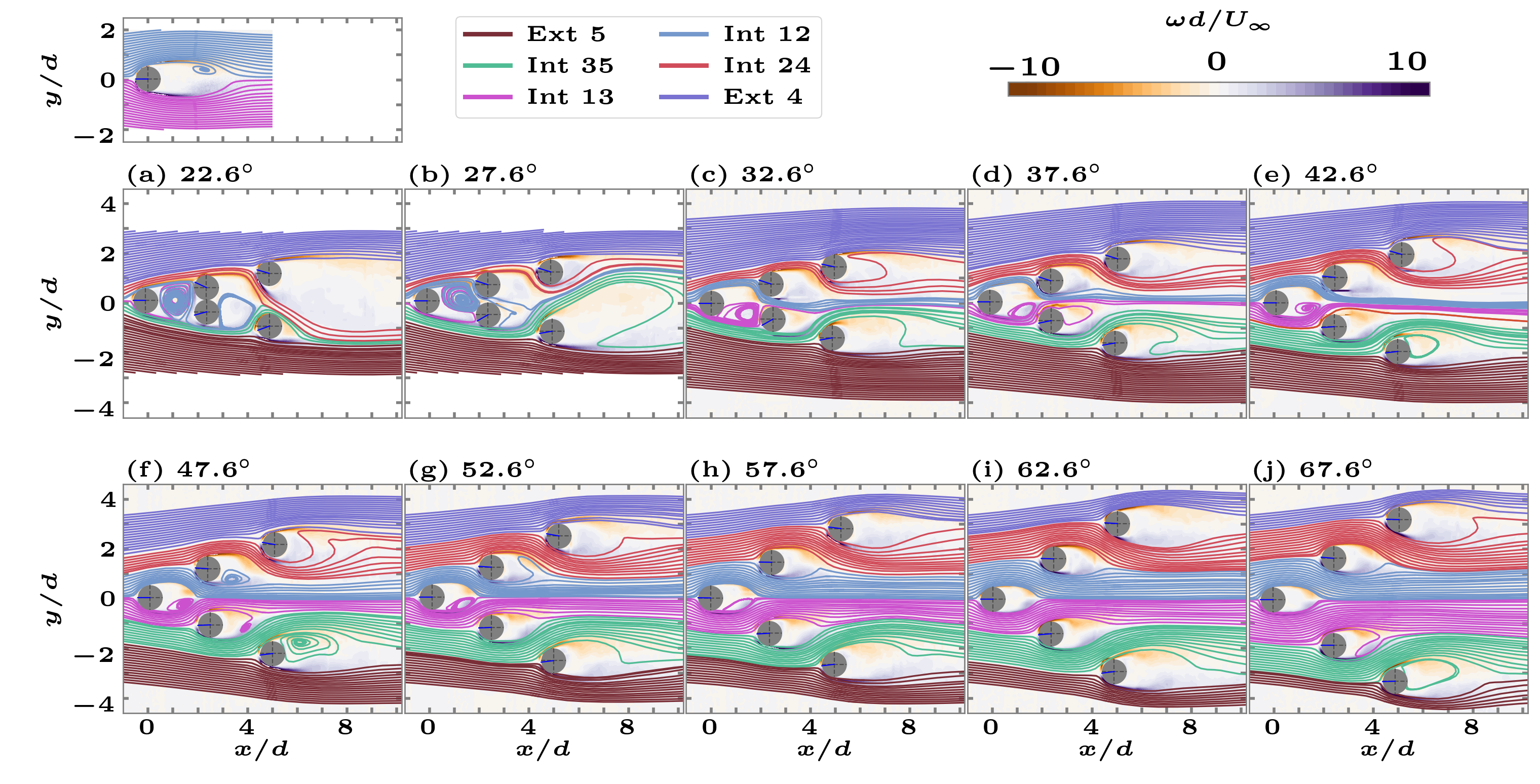}
    \caption{Streamlines of the flow for formations with $\phi$ going from $22.6^{\circ}$ to $67.6^{\circ}$ in steps of $5^{\circ}$. The streamlines are calculated via integration of the velocity fields starting from points along a vertical line at $x/d = -1$ and are color coded based on their trajectory around and in-between the different members. The dark blue lines inside the cylinders, connecting the centers of the cylinders and their circumferences, denote the location of the frontal stagnation points. The streamlines are overlaid on the mean vorticity contours, also shown in Fig.~\ref{fig:Omega} in the Supplemental Material for clarity. The streamlines for the solo cylinder is presented on the top as reference.}
    \label{fig:streamlines}
\end{figure}

To further explore the mechanism of variations in the drag force of each member ($C_{D,i}$) compared to the solo cylinder, presented in Sec.~\ref{sec:drag}, we focus on the velocity field past the formations from both global (this section) and local (upcoming Sec.~\ref{sec:local}) points of view. Figs.~\ref{fig:U} and \ref{fig:V} present the contours of the normalized mean streamwise and normal velocities in and around the formations, $u(x,y)/U_{\infty}$ and $v(x,y)/U_{\infty}$ respectively. For comparison, the velocity contours of the solo cylinder case are presented above the sub-figures. Figs.~\ref{fig:U}(a-e) and \ref{fig:V}(a-e) correspond to all formations with frontal overlaps among the members, and Figs.~\ref{fig:U}(f-j) and \ref{fig:V}(f-j) represent all the formations without frontal overlaps. The magnitudes and inward/outward direction of the bleeding flow between members of each echelon of the formations (with respect to the formation) are shown on the sub-figures of Fig.~\ref{fig:U} as well. Bleeding flow between members of the same row are shown in Tbl.~\ref{tab:bleed} in the Supplemental Material. To accompany the contour-plots and highlight the importance of bleeding flows, we also present the distribution of the streamlines of the mean flow calculated via integration starting from a point along the global $x = -d$ line upstream of the leading member as shown in Fig.~\ref{fig:streamlines}.

To further visualize the flow patterns, the streamlines of Fig.~\ref{fig:streamlines} are color-coded according to their trajectories throughout the flow field, with each separate color marking a group of streamlines passing in between or on the side of a specific member and their association with the bleeding or external flow. Thus, the streamline groups are separated according to the position of the stagnation points on the members as the point where the streamlines on each side divide into separate streams on different sides of the body. The upper- and lower-most groups are the streamlines passing on the exterior of members 4 and 5, and denoted by Ext 4 and Ext 5. The two middle groups, are the streamlines that pass in the interior in between members 2 and 4 on the upper echelon, called Int 24 (constituting the $\boldsymbol{u}^b_{24}$ bleeding flow), and members 3 and 5 on the lower echelon, called Int 35 (constituting the $\boldsymbol{u}^b_{35}$ bleeding flow). Similarly, the two streamline groups in the interior and between members 1 and 2 on the upper echelon, are called Int 12 ($\boldsymbol{u}^b_{12}$ bleeding flow) and 1 and 3 on the lower echelon, are called Int 13 ($\boldsymbol{u}^b_{13}$ bleeding flow). For the reference solo cylinder, the streamlines on either side of the cylinder are marked with the colors associated with Int 12 and Int 13 due to large similarity with member 1 of all the formations. 

\begin{figure}[!ht]
    \centering
    \includegraphics[width=0.98\textwidth]{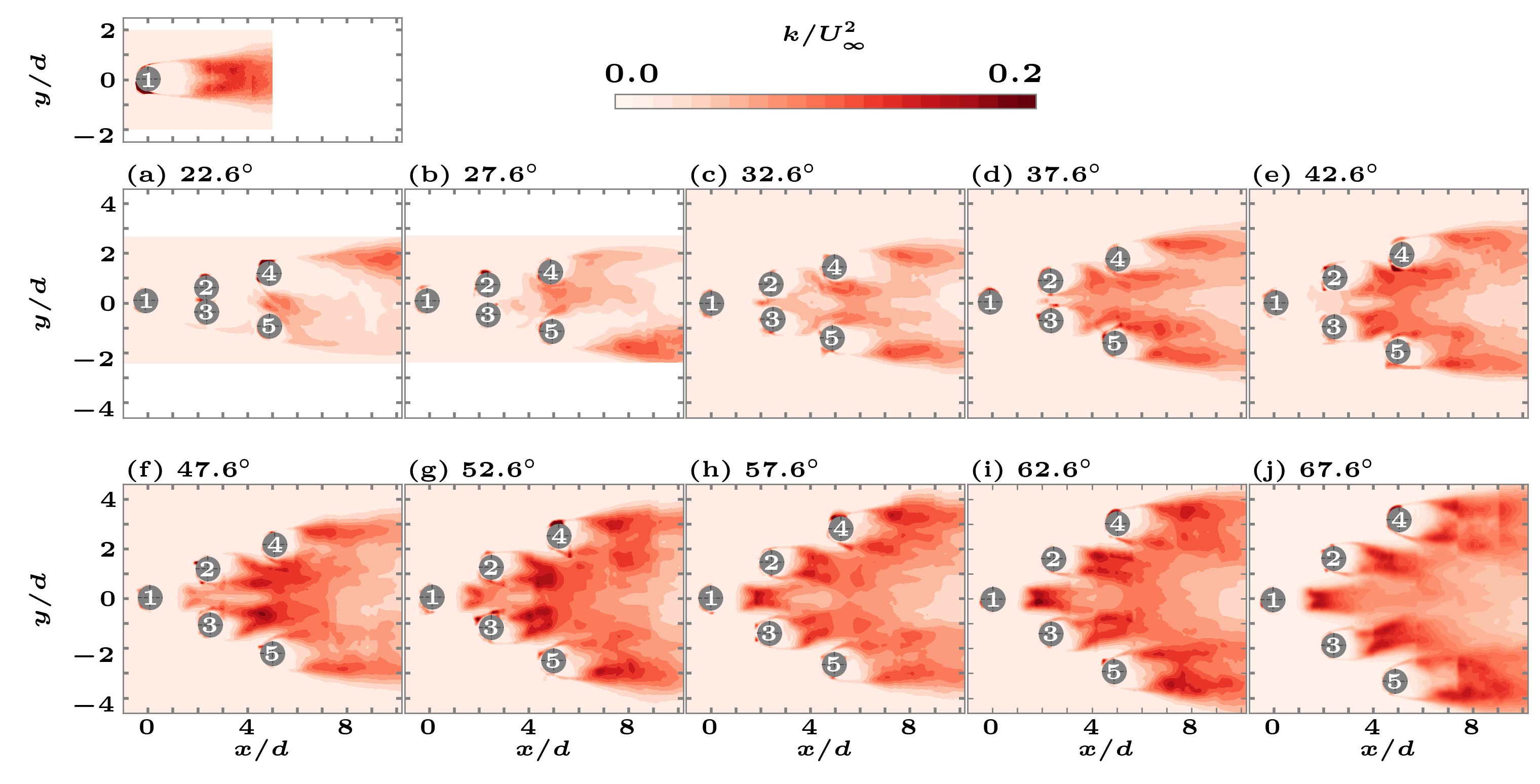}
    \caption{Contours of normalized turbulent kinetic energy, $k/U_{\infty}^2$, with $\phi$ going from $22.6^{\circ}$ to $67.6^{\circ}$ in steps of $5^{\circ}$. The turbulent kinetic energy contours for the solo cylinder is presented on the top as reference.}\label{fig:tke}
\end{figure}

Overall, we observe complex interactions between the wakes and the bodies of the members, where the characteristics of these interactions change as the angle of the $\pmb \vee$ is increased. As $\phi$ increases, the flow around the individual members of the formation slowly begin to resemble one another, especially in the case of the trailing members 4 and 5. However, in all the formations, the velocity fields around each of the members show differences compared with the solo cylinder case. Even at the largest allowed separation here (largest $\pmb \vee$ angle of $\phi = 67.6^{\circ}$), the wakes and the cylinders still sense the impact of their neighbors and do not reach a point where they are ``far away'' enough from each other for the flow to behave like flow past five isolated cylinders. 

The recorded differences, similar to the variations in the drag force discussed in Sec.~\ref{sec:drag}, are dependent on both the angle of the formation and the position of the member within the formation. On the one hand, the proximity of the bodies, and the choice of the streamwise distance (to be slightly lower than the vortex formation length), results in downstream bodies experiencing the impact of the wakes of the upstream  members. On the other, nonlinearities of the governing physics of the flow result in higher complexities of interactions that take place between the flow field past multiple members compared with the solo cylinder case. 

With consideration for the velocity contours and streamlines, i.e. flow dynamics, we can separate the studied cases into 3 groups, following a few distinct patterns observed in the flow: the first group consists of the two lowest angles with overlaps, $\phi=22.6^{\circ}$, and $27.6^{\circ}$, where the wake of member 1 encompasses members 2 and 3 and the wakes of the interior members are largely asymmetric about the $y=0$ line; the second group covers the cases of $\phi = 32.6^{\circ}$, $37.6^{\circ}$, $42.6^{\circ}$, and $47.6^{\circ}$ where the separation bubble in the wake of member 1 consists of two vortices fully enclosed by the streamlines of the bleeding flows $\boldsymbol{u}^b_{12}$ and $\boldsymbol{u}^b_{13}$ and the wakes of the rest of the members, especially members 4 and 5 are affected by the presence of the rest of the members; and the last group with $\phi \geqslant 52.6^{\circ}$, where the shape of the wake of the same member stays nearly similar as the angle is changed. We use these qualitative groups to focus on the physical mechanism of the flow field in the upcoming subsections. 

At the Reynolds number of the experiments here, the wake of a single cylinder reference case becomes turbulent and non-zero levels of turbulent kinetic energy, $k = 0.5(\overline{u'u'} + \overline{v'v'})$ is recorded in the wake with the maximum taking place around the formation length of $2.6d$. Therefore, here, in addition to the velocity contours and streamlines, we present the normalized turbulent kinetic energy, $k/U_{\infty}^2$, of the $\pmb \vee$-formations in Fig.~\ref{fig:tke}. In the overlapping formations, and especially lower angles, the slower flow in the interior of the array prohibits the flow from transitioning to turbulence  (compared with the wake behind an isolated cylinder) and therefore we see lower levels of $k$ in the upstream and in the interior of the formations (Figs.~\ref{fig:tke}(a-e)). However, significant wake-wake and wake-cylinder interactions as well as faster bleeding flows between the middle and trailing rows, lead to higher levels of turbulent kinetic energy in the downstream portion of these formations. As $\phi$ increases to non-overlapping cases, there is faster flow inside the array and more space is available for the wakes to become turbulent, resulting in wakes of all the formation members recording similar levels of turbulent kinetic energy as that of an isolated cylinder.

Below, we describe the key characteristics of the formations by dividing them into 3 groups with the most similar patterns based on the experimental observations as discussed earlier: 

\subsubsection{Most overlapping formations, $\phi=22.6^{\circ}$ and $27.6^{\circ}$} \label{sec:group1}

The leading members of formations with $\phi=22.6^{\circ}$, and $27.6^{\circ}$, shown in Figs.~\ref{fig:U}(a-b) and \ref{fig:V}(a-b), experience wakes that partially encompass members 2 and 3, leading to wider wakes in the normal ($y$) direction compared with the solo case. These 2 cases, with the lowest angles, only allow a small amount of flow in the interior of the formation, and the magnitudes of the equivalent bleeding flows, $U_{12}^b$ and $U_{13}^b$, are limited to less than about $0.09 U_{\infty}$. In parallel, the normal velocity in between members 1-2-3 shows larger magnitudes than what is seen in the wake of the solo cylinder. Especially for the formation with $\phi=22.6^{\circ}$, where members 2 and 3 are in contact, the wake mainly consists of one extremely slow circulation (vortical) zone in the separation bubble that is contained by the 3 bodies (as shown in Fig.~\ref{fig:streamlines}(a)), with the bleeding flow $\boldsymbol{u}^b_{12}$ flowing into space and the bleeding flow $\boldsymbol{u}^b_{13}$ flowing out.

\begin{figure}[!ht]
    \centering
    \includegraphics[width=1\linewidth]{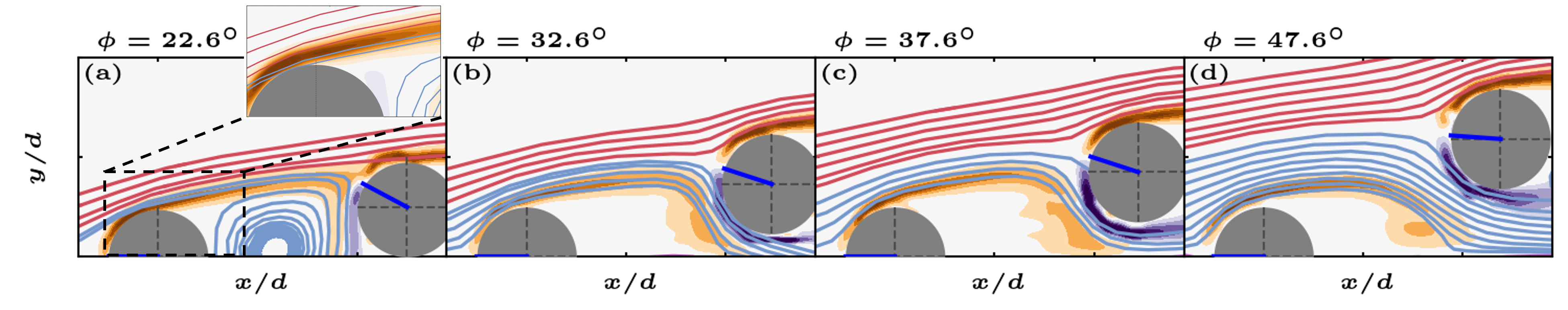}
    \caption{Intersection of shear layers (colored vorticity contours) with the streamlines Int 12 and Int 24 for formations with (a) $\phi = 22.6^{\circ}$, (b) $\phi = 32.6^{\circ}$, (c) $\phi = 37.6^{\circ}$, and (d) $\phi = 47.6^{\circ}$, zoomed in from Fig.~\ref{fig:streamlines}.} 
    \label{fig:streamline_shearLayer_combo}
\end{figure}

As seen from the color-coded streamlines in Fig.~\ref{fig:streamlines}(a), for formation angle $\phi = 22.6^{\circ}$, the limited physical separation between members 1-2-3 (especially between members 2-3) significantly regulates the bleeding flows. This is also clear from the limited extent of the streamlines Int 12 and Int 13  in the normal ($y$) direction. These two streamline groups, Int 12 and Int 13, are within the shear layers that are detached from either external sides of member 1, but they do not fully cover the entire thickness of the shear layer and a portion of the streamline groups Int 24 and Int 35 fill the rest of the shear layers' thickness as shown in Fig.~\ref{fig:streamline_shearLayer_combo}(a) and the inset, where Int 24 streamlines pass through the shear layer as marked with the faded orange region of the vorticity contour. The detached shear flow from exterior sides of member 1 is then divided into two streams at the stagnation points of members 2 and 3. A similar pattern is also seen for $\phi = 27.6^{\circ}$ but this time the thickness of the detached shear layers from the exterior of member 1 is mainly captured by both Int 12 and Int 13 and only a marginally small portion of the Int 24/Int 35.

In the $\phi = 22.6^{\circ}$ formation, on the one hand, streamlines Int 12 enter the space between 1-2-3 from the opening between members 1 and 2, turning into a separation bubble with one recirculating region, and then exiting from the opening between members 1 and 3, then again entering the interior of the formations through the space between members 3 and 5 and turning into another recirculation zone in the separation bubble behind member 3. From the opening between members 1 and 3, we also have streamlines of Int 13 entering the space and getting contained in the separation bubble. On the other hand, the portion of the shear layer (divided at the stagnation points), which coincides with portion of the streamlines of Int 35 and Int 24, move on the external sides of members 2 and 3. These streams then turn into the shear layers that detach from the exterior sides of these two members (2 and 3; follow Int 24 and Int 35 streamlines in Fig.~\ref{fig:streamlines}(a)). This pattern is partially similar to the ``shear flow reattachment'' phenomenon previously studied for the case of two cylinders in staggered configurations, with center-to-center distances less than 3 diameters and side-angles less than 10-20$^{\circ}$ \cite{sumner2000flow}. While on the external side of member 2, Int 24 turns into the shear layer that detaches from the wall, on the external side of member 3, both Int 12 (that has left the space between 1-2-3) and Int 35 together become the shear layer detaching from exterior of member 3. Therefore, only in the small region of the external attached shear layers, members 2 and 3 are in the vicinity of a fast moving flow (on the order of $U_{\infty}$) leading to higher levels of velocity gradient and shear stress at the wall. The rest of the perimeter of the cylinders are close to slower flow velocities (compared with $U_{\infty}$), leading to lower shear stresses. This results in the surface of cylinders 2 and 3, on average, experiencing lower levels of shear stress compared with a single cylinder. However, due to the asymmetry of the flow, member 3 is mostly encapsulated by Int 12 streamlines compared with member 2 that is in the vicinity of faster shear layers (Int 24).  

Between members 2 and 3 of the $\phi = 27.6^{\circ}$ formation, the bleeding flow $\boldsymbol{u}_{23}^b$ forces the flow to be more attached to member 3 than member 2 and leading to a larger wake and separation bubble behind member 2. This results in a larger pressure contribution to the drag of member 2, leading to uneven drag coefficients between members 2 and 3 of this formation with member 3 experiencing a lower drag force compared with the member 2 in the $\phi = 27.6^{\circ}$ formation. In the $\phi = 27.6^{\circ}$ formation, Int 24 streamlines are at the edge of the shear layers where the vorticity is approaching zero and this small portion of the shear layer of member 1 gets re-attached to members 2 and 3. As the angle of $\pmb{\vee}$-formation increases, the separation between members 1 and 2 opens the path for the bleeding flows $\boldsymbol{u}^b_{12}$ and $\boldsymbol{u}^b_{13}$ to increase as can be seen from the increase in the magnitudes of $U_{12}^b$ and $U_{13}^b$. When there is a slight gap between members 2 and 3, as for the case of $\phi = 27.6^{\circ}$ shown in Figs.~\ref{fig:U}(b) and \ref{fig:V}(b), the equivalent bleeding flows $U_{12}^b$ and $U_{13}^b$ are substantially higher ($\sim 9\%$) with the two bleeding flows becoming $\boldsymbol{u}^b_{23}$. However, the limited space between members 2 and 3 cannot change the pattern of the flow field (compared with $\phi = 22.6^{\circ}$) and a single recirculating vortex remains in the separation bubble between cylinders 1-2-3 (Fig.~\ref{fig:streamlines}(b)). Here again, we see the streamlines of Int 12 enter the separation bubble, create the recirculating region, and this time they leave the space via the gap between members 2 and 3. The Int 13 streamlines again enter the bubble and are confined in the recirculation. The Int 12 streamlines then leave and join the recirculating region in the separation bubble behind member 3. 

In the $\phi=27.6^{\circ}$ formation, however, the streamlines of flow stay attached to member 3 for a larger portion of the perimeter (Int 12 from one side, Int 13 from another, and Int 35 from the exterior side) compared with the flow detaching from member 2 as seen in Fig.~\ref{fig:streamlines}(b). The flow detachment in the downstream of member 2 results in a larger contribution of pressure drag compared with member 3 which results in the large disparity in the $C_{D,2}$ and $C_{D,3}$ as shown in Fig.~\ref{fig:CD_t_i}(a). Due to the described asymmetry, for both cases of $\phi = 22.6^{\circ}$ and $27.6^{\circ}$, member 2 of both formations see a well-defined, frontal stagnation point than member 3. This results in member 2 experiencing a greater momentum transfer from the fluid and higher drag force than member 3, as seen in Fig.~\ref{fig:CD_t_i}. Due to this asymmetric flow pattern, it is inevitable that the drag coefficients of the members of the same row (2,3 or 4,5) are different.

Going to members 4 and 5, with more space available between them in the normal ($y$) direction (compared with members 2 and 3), we see slightly larger bleeding flows, as shown with the equivalent bleeding values of $U_{24}^b$ and $U_{35}^b$ flowing into the interior of the formations (still less than $0.27 U_{\infty}$). These two bleeding flows are the main streams creating the $\boldsymbol{u}^b_{45}$ bleeding flow in between members 4 and 5. However, with the asymmetry of the entire flow field and the difference in the magnitudes of $U_{24}^b$ and $U_{35}^b$, we can see that the direction of the $\boldsymbol{u}^b_{45}$ is mainly dictated by the bleeding flow with the larger momentum, resulting in a large disparity between the wakes of the members 4 and 5 for these two formations. This can be better seen in Fig.~\ref{fig:V}(a) and \ref{fig:streamlines}(a) where a larger region of negative normal velocity is present past member 4 and 5 of $\phi = 22.6^{\circ}$ formation. We also see this in the pattern of the streamlines that encompass member 5 of the $\phi=22.6^{\circ}$ formation and moving away from member 4. The reverse of this pattern is seen for member 4 of the $\phi=27.6^{\circ}$, where as shown in Fig.~\ref{fig:V}(b) and \ref{fig:streamlines}(b), we see a large region of positive normal velocity past member 5 and in the wake of member 4, with the streamlines forcing the wake of member 4 to be smaller. In other words, between $U_{24}^b$ and $U_{35}^b$, the bleeding flow with the larger velocity magnitude (and thus momentum) dictates the distribution of the shear layers around the bodies. This results in a more detached flow for the body next to this streamline and forces the other streamline to bend and get more attached to the other member of the same row. For example, with $U_{24}^b>U_{35}^b$ in the $\phi = 22.6^{\circ}$ formation, we see a more detached flow for member 4 and more attached flow for member 5, resulting in member 5 experiencing a pressure recovery downstream and thus a lower drag force than member 4. An opposite situation happens for the $ \phi = 27.6^{\circ}$ formation where $U_{35}^b>U_{24}^b$ resulting in a more attached flow for member 4 and more detached flow around member 5, and member 4 experiencing a pressure recovery downstream and thus a lower drag force than member 5 (See Sec.~\ref{sec:local} for more on pressure recovery).

The limited space between members 1-2-3 for these 2 formations also impacts the turbulence statistics of the wake. At ${\rm Re}_d =1,100$ the wake of a solo cylinder turns turbulent with a formation length of around $2.6 d$ as shown in Fig.~\ref{fig:tke}. But for members 1, 2, and 3 of these 2 formations, the turbulent kinetic energy of the wake remains near zero and only in the wakes of the trailing members traces of non-zero turbulent kinetic energy is visible. 

Lastly, the asymmetry in the flow about the $y=0$ line or between members of the same row, especially 2 against 3, and 4 against 5 of the 5-cylinder system, has partial similarities to the phenomenon of three-cylinder fluidic pinball systems (3 cylinders of same diameter forming a $\pmb \vee$-formation as an equilateral triangle), which have been shown to have a bi-stable behavior undergoing a pitchfork bifurcation when the center-to-center spacing between the cylinders is below $\sim 2.3$ times the diameter of a single cylinder \citep{Bansal_2017,Lam_1988,Noack_2016,Deng_2020,Deng_2021}. As a result of this bi-stable behavior, the bleeding flow between the trailing cylinders doesn't stay symmetric and is deflected toward one of the trailing cylinders (more on the sensitivity analysis of this symmetry-breaking can be found in \citet{Noack_2016} and \citet{Deng_2020,Deng_2021}). Even in flows passed side-by-side cylinders, such biased or asymmetric flow fields with two wakes of different sizes is seen when side-by-side separations between 1.1-2.2 diameters are used (commonly referred to as intermediate separation) \cite{sumner2010two, sumner2000flow, zhou2016wake}. In all cases reported, the bleeding (or gap) flow is deflected and until the separation is set beyond 2.2 times the diameter, the asymmetric flow remains persistent. While the addition of the rest of the members impacts the overall dynamics of the flow making the wake-wake interactions more complicated, the similarities with the fluidic pinball and the side-by-side cylinder experiments merit future statistical investigation to study the stability and sensitivity of the flow past the formations with smaller $\pmb \vee$ angles and is out of scope of the current paper.

\subsubsection{Intermediate overlaps, $\phi=32.6^{\circ}-47.6^{\circ}$} \label{sec:group2}

For the rest of the formations with overlaps, namely $\phi = 32.6^{\circ}$, $37.6^{\circ}$, and $42.6^{\circ}$, and the marginal case of $\phi=47.6^{\circ}$, as the space between members 2 and 3 increases, the bleeding flow into the interior of the formation also increases. Unlike the bleeding flow $\boldsymbol{u}^b_{23}$ of $\phi=22.6^{\circ}$ and $27.6^{\circ}$ which is either non-existent or very small, starting from formation with $\phi = 32.6^{\circ}$ the bleeding flow between members 2 and 3 slowly acts as a means of establishing symmetry between the upper and lower echelons of the formation (Figs.~\ref{fig:U}(c-f), \ref{fig:V}(c-f), and \ref{fig:streamlines}(c-f)) and the differences between flow patterns around members 2 and 3, and then members 4 and 5 slowly disappears as the formation angle is increased. In the wake of member 1, two recirculating regions (vortices) form in the separation bubble between members 1-2-3 which are fully enclosed by the streamline groups of Int 12 and Int 13 (Figs.~\ref{fig:U}(c-f), \ref{fig:V}(c-f), and \ref{fig:streamlines}(c-f)). These two streamline groups flow into the interior of the formations bending inward at the stagnation points on members 2 and 3. In the case of $\phi = 32.6^{\circ}$, the wake of member 1 pushes the two stagnation points to be at non-negligible shifts with respect to the stagnation point of the solo cylinder (see the blue lines inside the cylinders in streamlines figures of Fig.~\ref{fig:streamlines}). Thus, the thickness of the wake of member 1 is still slightly larger than the solo case and certainly larger than the cases with no overlap (upcoming Sec.~\ref{sec:group3}). As the formation angle is increased and with larger space available between members 2 and 3 for the bleeding flows, streamlines Int 12 and Int 13 force the wake of member 1 to be slightly smaller in width and the stagnation points of members 2 and 3 slowly move back to the location similar to that of the single cylinder reference. These two streams also limit the horizontal extent of the wake to slightly less than $2.5d$ (forced by the second row of the members) compared with close to $3.5$-$4d$ for the solo cylinder. 

Similar to the most overlapping formations (discussed in Sec.~\ref{sec:group1}), the two streamline groups Int 12 and Int 13 are both part of the two shear layers (Fig.~\ref{fig:streamline_shearLayer_combo}(b,c)) that get detached from member 1 and at the stagnation points of member 2 and 3, and they bend inward turning into bleeding flows $\boldsymbol{u}^b_{12}$ and $\boldsymbol{u}^b_{13}$ into the interior of the formation. However, unlike the most overlapping formations, streamline groups Int 24 and Int 35 can mainly be placed outside of the boundary layers formed (on the sides of member 1). Therefore, for the formations of this group, Int 24 and Int 35 are only part of the shear layer that detaches from members 2 and 3 (Fig.~\ref{fig:streamline_shearLayer_combo}(b,c)) and with the shift in the stagnation points of these two and clear separations happening close to $90^{\circ}$ angles, these streamlines only stay attached to members 2 and 3 for less than or close to 25\% of the circumference of the cylinders on the exterior. On the other hand, the shear layers detached from member 1 (follow the trajectory of Int 12 and Int 13,  in Fig.~\ref{fig:streamlines}(c-f) and Fig.~\ref{fig:streamline_shearLayer_combo}(b,c)), reattach to the interior side of members 2 and 3, and due to the limited space between members 2 and 3, these two streams stay attached for slightly larger extents of the inner circumference of the cylinders. With the shear layer on the exterior of members 2 and 3 having faster velocities outside the boundary layer (comparable to that of the solo cylinder case), and the shear layers on the interior of the formation having slower speeds, members 2 and 3 of these formations, on average, experience lower levels of shear stress compared with the solo cylinder and this partially helps these members experience lower drag forces compared with the solo reference. As the angle of the formation increases, due to the increase in the magnitude of the bleeding flows, we then expect the shear stress on the interior to increase, thus increasing the drag force experienced by the same member (as seen in Sec.~\ref{sec:drag} and Fig.~\ref{fig:CD_t_i}). 

For formations with intermediate overlaps, as the formation angle is increased the flow field slowly takes a more symmetric pattern about the $y=0$ line. Cases with $\phi = 32.6^{\circ}$ and $37.6^{\circ}$ maintain the largest level of asymmetry between the two echelons (between member 2 and 3, and then member 4 and 5) as is seen both from the drag coefficients (Sec.~\ref{sec:drag}) and the wake contours and streamlines (Figs.~\ref{fig:U}(c-d), \ref{fig:V}(c-d), and \ref{fig:streamlines}(c-d)). Focusing on members 2 and 3, and looking at the streamlines around each of the bodies, we see the asymmetry leading to the flow field to be more attached to one of the cylinders and more detached from the other one. For example, for both $\phi = 32.6^{\circ}$ and $\phi = 37.6^{\circ}$ formations, member 3 experiences a more attached flow (see how the Int 13 encloses the body). This directly results in less pressure contributions to the drag of member 3 compared to member 2, resulting in a lower total drag force experienced by cylinder 3 (cumulative with the friction drag similar to the discussion in the previous Sec.~\ref{sec:group1}). For both these cases with the magnitude of the $U_{23}^b$ bleeding flow increasing (see Tbl.~\ref{tab:bleed} in the Supplemental Material), the drag force is larger than the lowest values recorded for $\phi = 22.6^{\circ}$ and $\phi = 27.6^{\circ}$. Starting from $\phi = 42.6^{\circ}$, the two echelons become more symmetric and the streamlines around members 2 and 3 of formations with $\phi = 42.6^{\circ}$ and $\phi = 47.6^{\circ}$ are nearly symmetric and similar and thus no clear difference is seen between the $C_D$ of members 2 and 3. 

The streamlines past members 4 and 5 that are moving into the formation (Int 24 and Int 35), have a similar pattern as streamlines Int 12 and Int 13 past members 2 and 3 moving into the formation. Streamline groups Int 24 and Int 35 are part of the detached shear layers off of members 2 and 3, which get reattached to members 4 and 5 on the inside by becoming part of the $\boldsymbol{u}^b_{45}$ bleeding flow, while streamline groups Ext 4 and Ext 5 are never part of the shear layers of the earlier members and only pass through the shear layers detached from members 4 and 5 on the exterior. These two streamline groups, beside the blockage caused by the group of bodies, do not experience any restrictions from outside and have the closest resemblance to the streamlines past the single member case. Ext 4 and Ext 5 streamline groups are only in contact with about 25\% (or less) of the circumference. On the other hand, Int 24 and Int 35 pass on the interior and both streams are faster than the streams past members 2 and 3 as the separation between the member is increased. Therefore, we expect slightly larger shear stress and friction drag for members 4 and 5 of the trailing row compared with members 2 and 3 in the middle row. In addition, following the streamlines in Figs.~\ref{fig:streamlines}(c-f), we see more attached streamlines around member 4 of the $\phi = 32.6^{\circ}$ (following Int 24) and member 5 of the $\phi = 37.6^{\circ}$, $\phi = 42.6^{\circ}$, and $\phi = 47.6^{\circ}$ (follow Int 35). This asymmetry, again, results in lower pressure contributions to the drag for these members compared to the other member in the same row. Therefore, member 4 of the  $\phi = 32.6^{\circ}$ has a lower drag force compared to member 5 in the same trailing row, and for $\phi = 37.6^{\circ}$, $\phi = 42.6^{\circ}$, and $\phi = 47.6^{\circ}$, we see the reverse, where member 5 records a lower drag force compared to member 4 of the same formation.

Lastly, the increase in the gap between the members, also results in the turbulent kinetic energy of the flow to increase in the wakes of the bodies in the interior of the formation, as we can see larger magnitudes in Figs.~\ref{fig:tke}(c-f), especially in the wakes of members 2 and 3. It is only in the case of $\phi = 47.6^{\circ}$ where we see non-zero kinetic energy levels in the wake of member 1 prior to the position of members 2 and 3, and from this angle forward, we record turbulent kinetic energy levels close to the maximum value reported in the wake of the solo cylinder case. While the maximum values of the turbulent kinetic energy are lower than that of the solo cylinder case, with the limitations in the space for the wakes of members 2 and 3 to develop, the formation lengths behind these cylinders are less than that of the solo cylinder case.

\subsubsection{Non-overlapping formations, $\phi=52.6^{\circ}-67.6^{\circ}$} \label{sec:group3}

For all the cases with $\phi \geqslant 52.6^{\circ}$ (no overlaps in the streamwise direction), we see the width of the wake of member 1 to be smaller than the previous two groups (see Figs.~\ref{fig:U}(g-j) and \ref{fig:V}(g-j)), as the space between members 2 and 3 allows for larger magnitudes of bleeding flow to move between the members. Right before members 2 and 3, the normal velocity records larger magnitudes (especially on the interior sides) compared with the solo cylinder case. Similarly, we can see in Fig.~\ref{fig:streamlines}(f-i) the streamlines on either side of each of the bodies divide upstream of the member at the stagnation points, go around the bodies, and then come back together, closing the bodies, their wakes, and the separation bubbles inside a fictitious boundary. For example, similar to the formations with intermediate overlaps, the streamline groups Int 12 and Int 13 completely encompass member 1, its wake, and separation bubble, before the centerline of members 2 and 3 (horizontal distance of less than $2.5d$). Similarly, for members 2 and 3, we see Int 12 and Int 24, as well as Int 13 and Int 35 separate at the stagnation points of members 2 and 3 and then the two streamline groups come together before members 4 and 5, fully enclosing members 2 and 3 and their wakes and separation bubbles. The physical size of these regions for members 1, 2, and 3, bounding each of the bodies and the wakes and separation bubbles, are markedly smaller than the boundary created by the streamline around the solo cylinder case, due to the continuity dictating the existence of the bleeding flows $\boldsymbol{u}^b_{23}$ and $\boldsymbol{u}^b_{45}$, between members 2-3 and 4-5. As a result, the streamlines are forced to be closer to the bodies, and bend and come back together at an earlier distance compared with the same extent in the single cylinder case. This in turn results in shear layers separating at a slightly later point along the circumference of the bodies at angles larger than $90^{\circ}$ compared to the usual $80-85^{\circ}$ separation points seen in the solo cylinder case here and in previous works \cite{Kundu_2008}. So, while this can slightly increase the frictional drag force, it mostly helps with the pressure drag on the body which ultimately results in the leading member of all the formations experiencing a lower drag force even until the largest $\pmb \vee$-angles and members 2 and 3 also seeing slight levels of drag reduction of between 5-8\% for $\phi = 52.6^{\circ}$ and $57.6^{\circ}$. Lastly, this limited length of the wake for members 1, 2, and 3 also impacts the development of the turbulent kinetic energy in the interior of the formation (see Fig.~\ref{fig:tke}(g-j)) and we see the vortex formation lengths (location with the largest turbulent kinetic energy \cite{bearman1965investigation, bearman1969vortex, bearman1984vortex, Williamson_1996}) to be less than that of the solo cylinder case. 

While the wake of member 1 is nearly symmetric about its horizontal centerline, the rest of the members 2-5 of these formation still experience asymmetries about their local horizontal centerlines ($y=y_c$) as the streams on the interior and exterior sides of the members have clear differences as seen in Fig.~\ref{fig:streamlines}(g-j). Additionally, members of the two echelons (2 and 4, against 3 and 5) inevitably see slight bits of asymmetries about the global horizontal direction ($y=0$ line). However, as the formation angle increases the symmetry of the upper and lower echelons of the formation about $y=0$ becomes stronger. These asymmetries are visible for members 2-5 in Fig.~\ref{fig:U}(g-j), where the wakes of the members have curved, asymmetric teardrop shapes. The asymmetry in the flow past either cylinder 2 or 3 is also clearly seen in the distribution of the normal velocities in Figs.~\ref{fig:V}(g-j); on the interior side of the members, the streamlines constituting the bleeding flows $\boldsymbol{u}^b_{23}$ go straight through toward the space between members 4 and 5 and join the bleeding flow $\boldsymbol{u}^b_{45}$, while the streamlines on the exterior of the formation bend inward and shape the bleeding flows $\boldsymbol{u}^b_{24}$ and $\boldsymbol{u}^b_{35}$.  The trailing members 4 and 5, with all the space available in their downstream, develop wakes with closer similarities to the wake of a solo cylinder (but with the aforementioned asymmetries). 

Even with the differences in the flow field and the streamlines, members 4 and 5 of these formations record similar drag values as that of the solo cylinder as seen in Fig.~\ref{fig:CD_t_i} in Sec.~\ref{sec:drag}. For members 4 and 5 in wider formations, with the larger distance between the two members of the same row, the bleeding flow has enough space to develop without forcing the streamlines to move closer to each other. Hence, with more given freedom, for $\phi\geqslant 57.6^{\circ}$ we see the drag coefficient of members 4 and 5 to return to that of the solo cylinder case and even member 5 recording slightly higher drag force than the solo cylinder. Overall, we see nearly similar drag forces recorded for majority of the members (interior and trailing) and especially for the case with $\phi = 67.6^{\circ}$, the drag force most of the members experience are close to the drag force of a single cylinder reference case.

\subsection{Local view of the flow field around each member}\label{sec:local}

To further compare the effect of the different formation angles on each of the cylinders, we change our point of view from the global coordinate system at the center of cylinder 1, to a local coordinate system attached to the center of each member (denoted by $(x_c, y_c)$), and present the velocity and pressure fields as seen by a specific member in the upstream and downstream positions compared to the same member in other formations (different $\phi$).

\begin{figure}[!ht]
    \centering
    \includegraphics[width=0.98\linewidth]{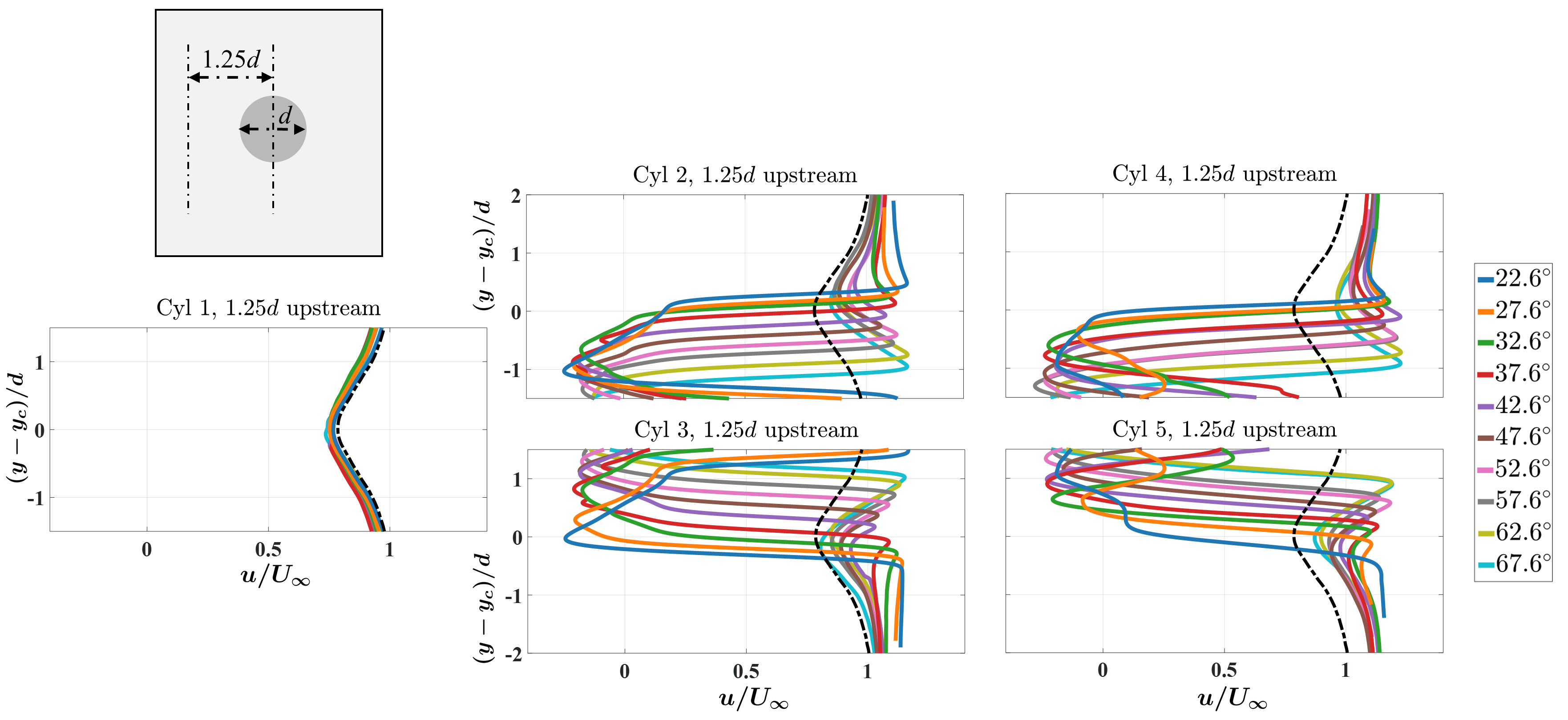}
    \caption{Streamwise velocity profiles, $u/U_{\infty}$, $1.25d$ upstream of the center of the members, plotted with respect to the center of the member as shown in the top left schematic. Black dash-dotted line represents the velocity at $x/d = 1.25$ upstream of a single cylinder.  }
    \label{fig:u_upstream}
\end{figure}

\begin{figure}[!ht]
    \centering
    \includegraphics[width=0.98\linewidth]{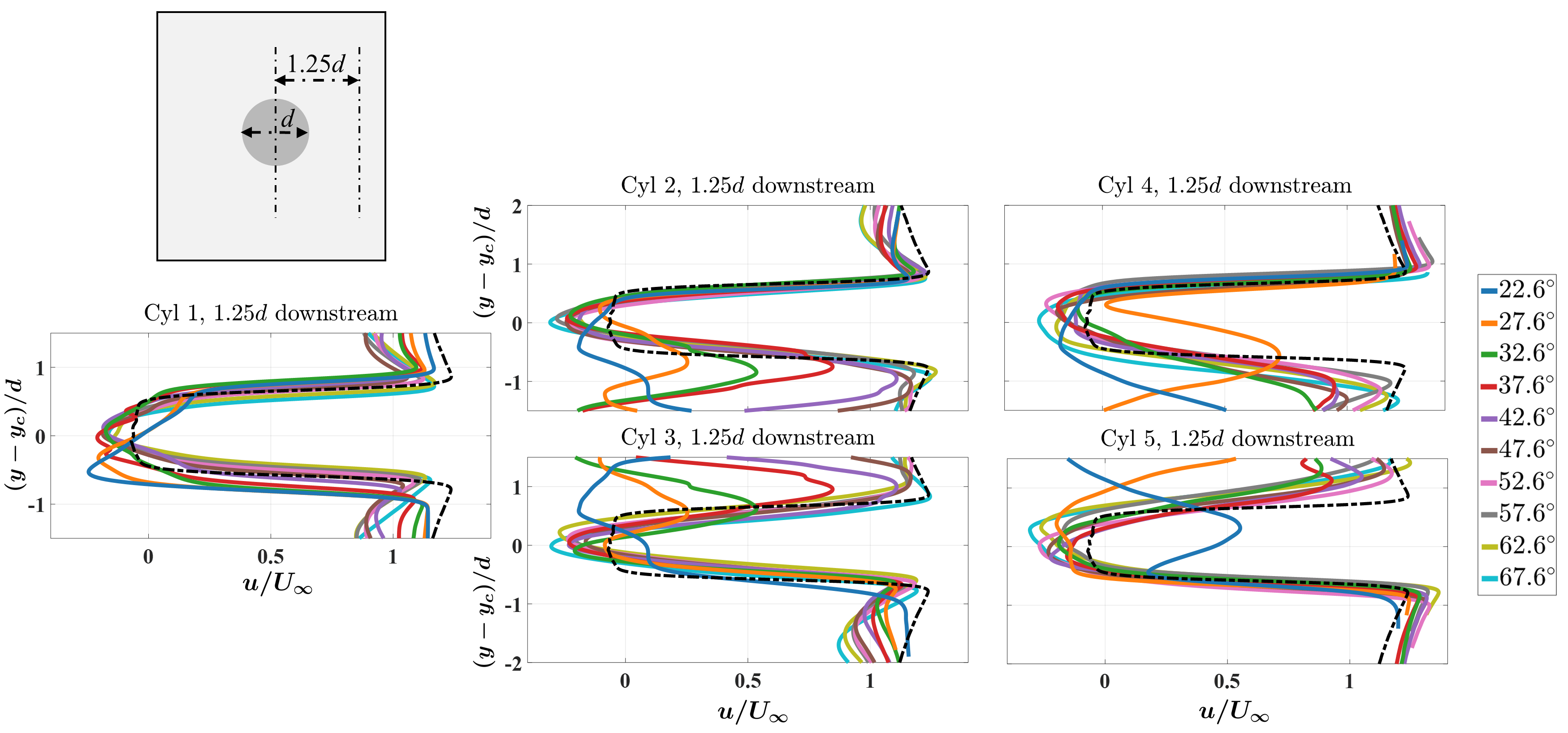}
    \caption{Streamwise velocity profiles, $u/U_{\infty}$, $1.25d$ downstream of the center of the members, plotted with respect to the center of the member as shown in the top left schematic. Black dash-dotted line represents the velocity at $x/d = 1.25$ downstream of a single cylinder.}
    \label{fig:u_downstream}
\end{figure}

The velocity profiles upstream of each of the members (Fig.~\ref{fig:u_upstream}) experience large variations as a function of the formation angle, than the profiles on the downstream side (Fig.~\ref{fig:u_downstream}). Among the different cylinders, for all the members of interior and trailing rows (2-5), velocity fields recorded at $1.25d$ upstream of each of the members (Fig.~\ref{fig:u_upstream}(b-e)) show large differences as the formation is changed and when compared with the single cylinder case (shown with black dashed lines). It is only the case of the leading member (Fig.~\mbox{\ref{fig:u_upstream}}(a)) where the upstream velocity profiles nearly collapse on each other. Due to the presence of multiple bodies in the path of the flow (as reported previously \mbox{\cite{suchandra2024}}), the upstream velocity of all the formations is slightly lower than that of the reference solo cylinder case. 

The upstream differences seen by the rest of the members (interior and trailing, 2-5) are a direct result of the variations in the angles of the formations and are mostly in the form of similar profile patterns that are at transverse shifts in the $y$ direction with respect to each other. As it has been discussed before \cite{suchandra2024}, within a formation of non-lifting bodies, three phenomena dictate the dynamics of the flow field around the members: (1) the wake of the upstream members, (2) the stagnation points of the downstream members, and (3) the bleeding flow passing in between two members. Focusing on the formation with the widest angle here, $\phi = 67.6^{\circ}$, one can clearly identify these three phenomena as three separate regions in the velocity profiles of cylinders 2-5. The wake deficit is seen as the largest slow-down region in the profile, bleeding flows correspond to regions of the speed up in the flow ($u/U_{\infty}>1$) which also limit the lateral extent of the wake in the $y$ direction, and a smaller region of slow-down (velocity magnitude) due to the impact of the upcoming body and stagnation point being sensed by the flow. These three regions are easily distinguishable in the rest of the non-overlapping cases with $\phi > \phi^*$, but as we reduce the formation angles, the separation between the three regions shrinks and the distinction between the bleeding flow and the slow down due to stagnation becomes less clear. Especially for overlapping cases, $\phi<\phi^*$, where the upstream wake region shows its impact within the $-0.5d<y-y_{\rm c}<0.5d$ (immediate frontal view), the velocity profile is mainly dominated by the wake deficit, and a more dilute effect of the bleeding flow and the slow-down from the upcoming stagnation points are visible. As it can be seen from the blue lines marking the location of stagnation points on each body in the streamline plots of Fig.~\ref{fig:streamlines}, as $\phi$ decreases, the location of the stagnation point is pushed outward with respect to the $180^{\circ}$ horizontal angle of the stagnation point of the solo cylinder case. For lager $\phi$ angles, we see the location of the stagnation point slowly returning back to the same position as that of the solo cylinder, but the flow past the members maintain an asymmetry due to the geometric constraint between two side-by-side members. This also contributes to the stagnation points being visible on the line plots as smaller slow-down regions, from upstream of the cylinders for all $\phi \geqslant 37.6^{\circ}$ formations, as seen in \mbox{Fig.~\ref{fig:u_upstream}}. For overlapping formations, such as $\phi = 42.6^{\circ}$, $\phi=37.6^{\circ}$, and $\phi = 32.6^{\circ}$, the slow-down visible due to the stagnation point follows the outward shift we discussed in Sec.~\ref{sec:group2}, and also seen in Fig.~\ref{fig:u_upstream}. For the two formations with $\phi = 22.6^{\circ}$ and $27.6^{\circ}$ the slow-down due to the stagnation point is much more dilute for members 2 and 3 and this is again due to the fact that the wake of member 1 encompasses members 2 and 3 as discussed in Sec.~\ref{sec:group1}. As the formation angle increases, the location of the slow down moves closer to the $y=y_c$ line, similar to that seen for the reference solo cylinder as shown with the black dashed line.

The wakes of the upstream members, play the largest role in dominating the velocity profiles seen upstream of the members, compared with the bleeding flow and the presence of the upcoming stagnation points (Fig.~\ref{fig:u_upstream}(b-e)). Comparing the velocity profiles upstream of the interior members 2 and 3 and the profiles of member 1 of all the formations, one can see how the impact of the wakes of the previous members change the velocity distribution compared to only having one upcoming stagnation point in the flow. The wakes of the previous members are always located directly behind the previous member. Increasing the formation angle, increases the transverse separation, $y$, of the valley of the velocity deficits with respect to the $y=y_c$ line, with the valleys shifting away toward the interior of the formation  with respect to the center of the upcoming member. Therefore, for non-overlapping $\phi>\phi^*$ formations, the extent of the upstream wake region moves away from the immediate view of the member ($-0.5d<y-y_{\rm c}<0.5d$ region). The immediate frontal view upstream is dominated by the slow-down from the upcoming stagnation point and the speed up due to the bleeding flows. As the formation angle is increased, the velocity profiles asymptotically lean toward a profile similar to that of a solo cylinder, but as seen in Sec.~\ref{sec:global}, the separations tested here are not yet large enough to reach the point where all the cylinders are uninformed of each other's presence. As expected, the exterior sides of the profiles resemble the profile of the solo cylinder more than the interior sides. This asymptotic behavior is clearer for the members of the interior row, 2 and 3. Members 4 and 5 of the trailing row experience a larger velocity in the exterior, mainly due to the large blockage caused by the entire formation. For the most-overlapping $\phi = 22.6^{\circ}$ and $27.6^{\circ}$ cases, a wide extent of the views of the members is covered by the upstream wakes as well as the streamlines of the bleeding flow that join the recirculating regions in the wakes. This results in the impact of the bleeding flow, the upcoming stagnation points, and the wake to be blended together.

The dominance of the impact of the wakes of the members on the flow field is better visible when focusing on the velocity profiles downstream of the members (Fig.~\ref{fig:u_downstream} - viewing from the downstream side of the members). Since the wakes are formed immediately downstream of the bodies, the velocity profiles seen by the same member in different formations, present closer similarities to each other, with all the wake deficits nearly collapsing on top of each other and the impacts of the bleeding flow and the stagnation point downstream are only seen in the form of the differences toward the interior of the formations. As discussed in Sec.~\ref{sec:global} and seen in the profiles of Fig.~\ref{fig:u_downstream}, in formations with frontal overlaps ($\phi<\phi^*$), the width of the wake deficit in the downstream of the leading member is larger than the wakes of formation with no overlaps ($\phi>\phi^*$). Due to the bleeding flows from either side of the cylinders, the valley of the wake deficits behind all the members present a sharper pattern compared with the blunt valley of the wake deficit of the solo cylinder. The bleeding flows alter the development of the wakes leading to earlier (spatial) development of recirculating regions in the wake and the separation bubble than that of the solo cylinder (see Figs.~\ref{fig:U}, \ref{fig:V}, and \ref{fig:streamlines}) where we see larger negative $u$ values in the wakes of formation members compared to the solo cylinder. Between all the cylinders, the members of the middle row, i.e. members 2 and 3, see a narrower 
wake compared with that of the trailing and leading members, independent of the formation angle. This can be mainly associated with the constraints of the bleeding flow imposed by the conservation of mass and the available geometric opening between the bodies. As a result, the streamline groups Int 24 and Int 35, passing on the sides of members 2 and 3 are forced to flow through the opening between members 4 and 5 and becoming part of the $\boldsymbol{u}^b_{45}$ bleeding flow. This is while members 4 and 5 have no downstream limitations. These streamlines thus bound the flow and limit the development of the wakes of members 2 and 3 resulting in a narrower wake compared to the wakes of members 4 and 5 that are somewhat (but not fully) free to develop downstream when compared with the single cylinder case. 

In case of the interior members 2 and 3, and in most formations with frontal overlaps ($\phi<42.6^{\circ}$) the separation between the side-by-side wake deficits of the two members is not sufficient for the magnitude of the bleeding flow ({seen along the $(x-x_c) = 1.25d$}) to be able to fully recover to the $U_{\infty}$ and this results in slight differences between the velocity fields as a function of the formation angles. Starting from $\phi = 42.6^{\circ}$, we see minimal variations between the velocity profiles seen by the same member in different formations. The same pattern is also seen for members 4 and 5 where only formations with $\phi \leqslant 27.6^{\circ}$ see a clear difference between the velocity profiles due to the bleeding flow and the upcoming stagnation points, and formations with $\phi > 27.6^{\circ}$ experience similar velocity profiles as the formation angle is changed. 

\begin{figure}
    \centering
    \includegraphics[width=0.98\linewidth]{Fig10.png}
    \caption{Pressure coefficient profiles, $C_p$, $1.25d$ upstream of the center of the members, plotted with respect to the center of the member. Black dash-dotted line represents the pressure coefficient at $x/d = 1.25$ upstream of a single cylinder.}
    \label{fig:cp_upstream}
\end{figure}

\begin{figure}
    \centering
    \includegraphics[width=0.98\linewidth]{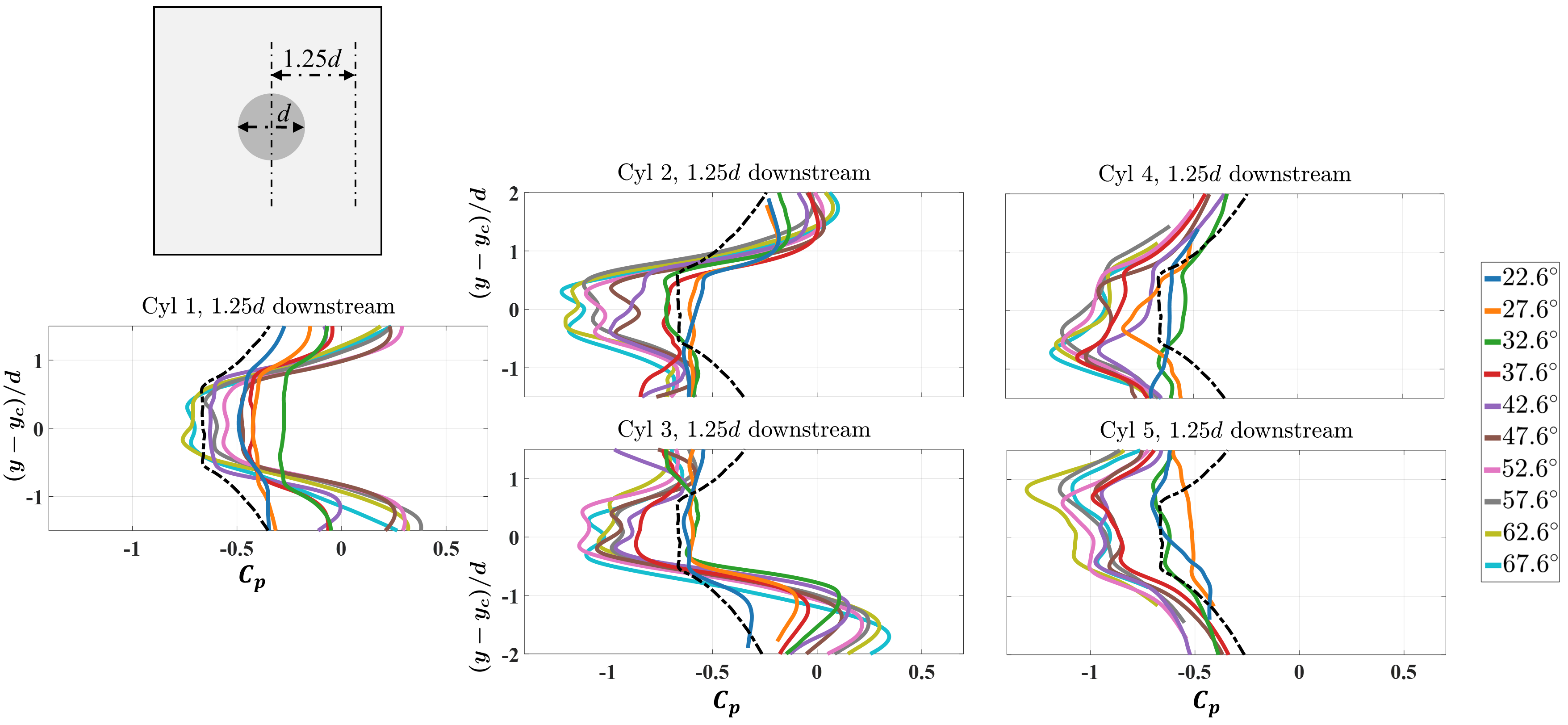}
    \caption{Pressure coefficient profiles, $C_p$, $1.25d$ downstream of the center of the members, plotted with respect to the center of the member. Black dash-dotted line represents the pressure coefficient at $x/d = 1.25$ downstream of a single cylinder.}
    \label{fig:cp_downstream}
\end{figure}

We also compare the distribution of the pressure coefficient, $C_p = (p - p_{\infty})/(0.5 \rho U_{\infty}^2)$, as seen locally by each member in the upstream and downstream positions, shown in Figs.~\ref{fig:cp_upstream} and \ref{fig:cp_downstream}. The key difference in the pressure coefficients (compared with a solo cylinder) is in the pressure recovery occurring in the downstream of the members of the formations, where, on average, the difference between the pressure distribution upstream and downstream of the members are lower than that of the single cylinder case. In case of the leading member, the presence of multiple bodies downstream of the flow results in a slightly larger pressure coefficient to be recorded upstream, while on the downstream, we see larger pressure coefficients at all points along the $x = 1.25 d$ line than that of the single cylinder case. For non-overlapping cases, while a pressure recovery is visible away from the $y=y_c$ line, the pressure coefficient in the wake is similar to that of the solo cylinder case. Overall, the pressure recovery, added to the impact of the wakes and the bleeding flows around member 1 leads to the leading member universally benefiting from its position in the formation. 

Moving to members 2 and 3, the pressure coefficient upstream of the overlapping cases with $\phi<\phi^*$ are visibly different from those of the non-overlapping cases, with non-overlapping ones having a profile similar to that of the solo cylinder on the external side of the formation at $y-y_c>0.5d$ for member 2, and at $y-y_c<-0.5d$ for member 3. For all the overlapping cases, and in the interior side of the non-overlapping formations, we should accept some level of pressure drop compared to the pressure distribution upstream of a solo cylinder cases. Here, the pressure coefficient follows the pattern dictated by the wake of member 1 which is clearly different from the flow field upstream of a single cylinder. However, as the formation angle increases, the pressure coefficients on the external sides of members 2 and 3 start to resemble that of a solo cylinder. Comparing the upstream and downstream, the pressure coefficients of overlapping formations, on average, experience a lower pressure drop compared with the non-overlapping cases. Especially for $\phi \leqslant 42.6^{\circ}$ formations, we see that the differences in the $C_p$ of the upstream and the wakes of members 2 and 3 show lower drops compared to the pressure drop seen between the upstream and downstream of a solo cylinder case. Since the key impact of the pressure on the drag of each member is in this difference between upstream and downstream, most of the overlapping cases, as well as $\phi = 47.6^{\circ}$, experience a benefit from the pressure recovery that is possible due to the presence of upcoming members and their stagnation points. For formations with the largest overlaps, namely, $\phi = 22.6^{\circ}$, $27.6^{\circ}$, and $32.6^{\circ}$, the pressure recovery, along with the flow attachment due to the complex bleeding flow patterns described in Sec.~\ref{sec:global}, result in the substantial levels of drag reduction recorded for members 2 and 3 of these formations.

As we move to the trailing row, on the upstream of members 4 and 5, the profile of the pressure coefficients is dominated by the impact of the wake of the members 2 and 3, and shows clear differences with the wake of a single member case (as expected). But on the downstream, the profiles of pressure coefficient show much closer similarity to that of the single cylinder case as there is no geometric restrictions to the development of the wake downstream. Similar to members 2 and 3 discussed in the previous paragraph, members 4 and 5 of overlapping formations experience a slightly larger pressure drop (between upstream and downstream positions) compared to the solo cylinder case. However, as discussed in Sec.~\ref{sec:group1}, the pressure recovery behind member 4 of the $22.6^{\circ}$ formation, and member 5 of the $27.6^{\circ}$ formation along with the larger flow attachment around these two members result in lower drag forces compared with their side-by-side members in the same row. For non-overlapping formations with $\phi \geqslant 57.6^{\circ}$ the pressure drop across members 4 and 5 are overall quite similar to that of the solo cylinder cases, which also leads to members 4 and 5 of these formations experiencing drag coefficients in the similar range of that of the solo cylinder, with member 5 even experiencing slightly larger values of drag as seen in Fig.~\ref{fig:CD_t_i} in Sec.~\ref{sec:drag}.

\subsection{Temporal characteristics and vortex shedding patterns} \label{sec:vortexshedding}

While the main focus of this work is on the drag force experienced by the bodies as a performance metric for individual members and the group, the asymmetry observed in the flow field leads one to also question the possibility of variations in the lift forces experienced by the bodies. A single cylinder in flow experiences a steady state drag and zero steady state lift, but the instantaneous vortex shedding pattern behind the cylinder results in a temporal distribution of lift that averages to zero in steady state conditions. Thus, cylinders are generally considered to be non-lifting. However, the presence of additional members in a group and the asymmetry seen in the flow field and especially pressure, as discussed in Sec.~\ref{sec:global} and \ref{sec:local}, raises the possibility of the cylinders experiencing mild levels of lift force. Using the control volume analysis of Sec.~\ref{sec:force_cal}, we can estimate the lift of bodies; for the cylinders experiencing the most asymmetry (like the ones in the middle and trailing rows of the most overlapping cases, $\phi = 22.6^{\circ}$ and $27.6^{\circ}$; see Figs. \ref{fig:U} and \ref{fig:V}), we record the largest lift-to-drag ratios, with the ensemble-averaged lift forces being $\sim 10-15 \%$ of the ensemble-averaged total net force based on CV calculations. However, it should be noted that due to the low magnitudes of the lift force captured here, a measurement technique with even higher resolution than what used here is required to capture this force (such as higher resolution load cells). 

\begin{figure}[!ht]
    \centering
    \includegraphics[width=0.998\linewidth]{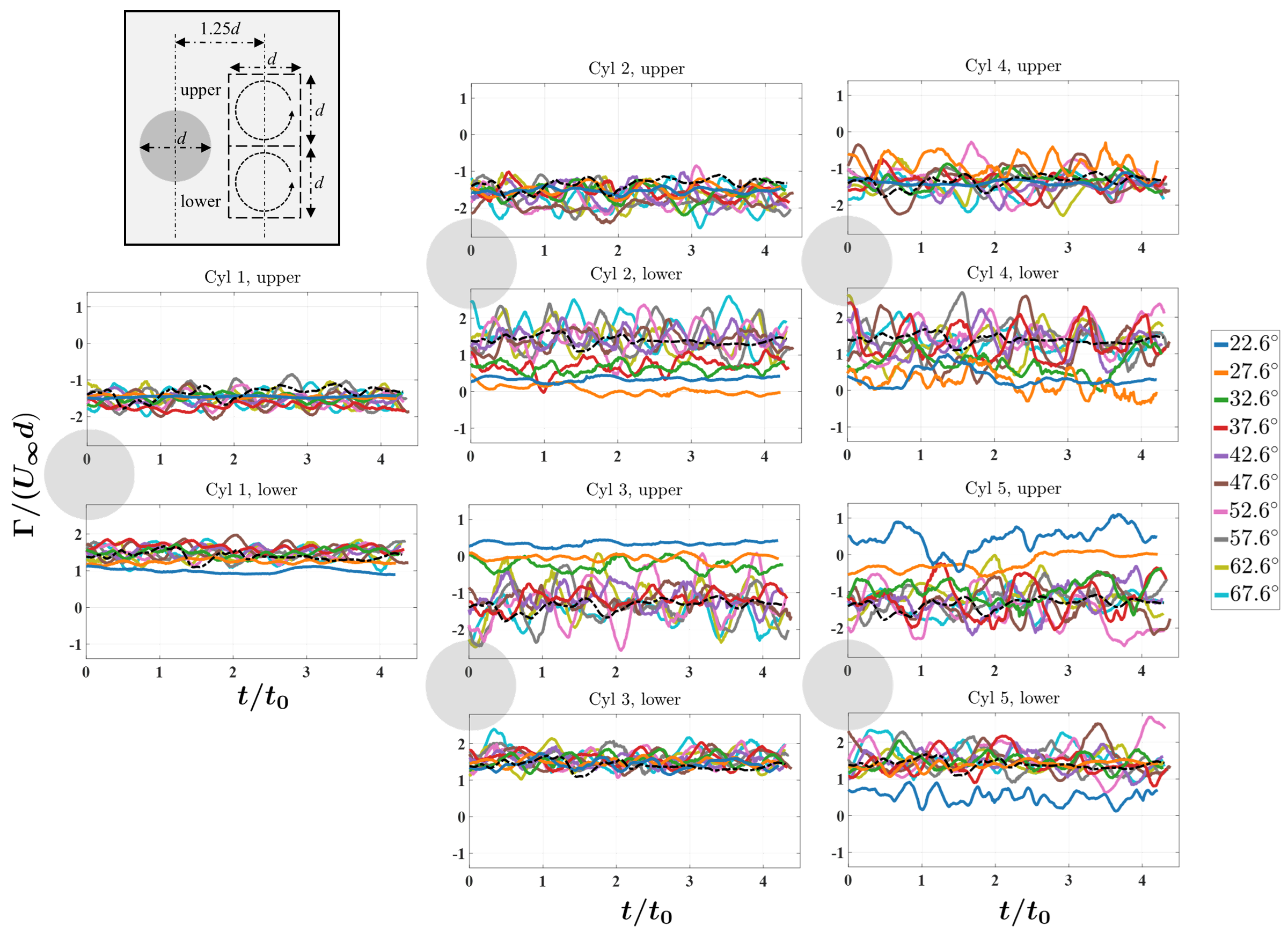}
    \caption{Normalized instantaneous circulation, $\Gamma / (U_{\infty} d)$, as a function of non-dimesnionalized time, $t / t_0$, calculated in a $d \times d$ square region $1.25d$ downstream of the upper and lower shoulders (center of the square regions are downstream of the uppermost and lowermost points of the cylinder) of each cylinder member for all the $\pmb \vee$-formations (these square regions are shown on top-left of the figure). For visualization purpose, we use light gray circles to qualitatively denote the relative position of the formation members and connect the two plots of the upper and lower circulations for the same member. $t_0$ denotes the vortex shedding time period for an ideal long circular cylinder. The black dashed-dotted lines correspond to the upper and lower circulations due to vortex shedding behind a single, isolated cylinder.}
    \label{fig:circ}
\end{figure}

Instead of focusing directly on the lift force, and knowing the impact of vortex shedding and the circulation on the lift, we explore the temporal characteristics of flow field around the members of $\pmb \vee$-formations, and calculate the circulation $\Gamma = \rcircleleftint \pmb{u} \cdot d \pmb{\ell}$ along two chosen closed curves encompassing the upper and lower regions in the wake of the members ($y>y_c$ and $y<y_c$) and present the results as a function of time $t$ in Fig.~\ref{fig:circ}. Schematics of these regions over which the integrals are carried out are shown in top-left corner of the Fig.~\ref{fig:circ}. The black dash-dotted lines in Fig.~\ref{fig:circ} correspond to the upper and lower circulations due to vortex shedding behind a single, isolated cylinder. $t_0 = 1/f_0$ denotes the vortex shedding time period for an ideal long circular cylinder, estimated from the literature on the Strouhal number (${\rm St}_d = d/(t_0 U_{\infty})$) variation with the Reynolds number (${\rm Re}_d$) \citep{Lienhard_1966}. For $\rm{Re}_d \approx 1100$ used in the current study, the corresponding Strouhal number is ${\rm{St}}_0 \approx 0.215$.

\begin{figure}[!ht]
    \centering
    \includegraphics[width=0.998\textwidth]{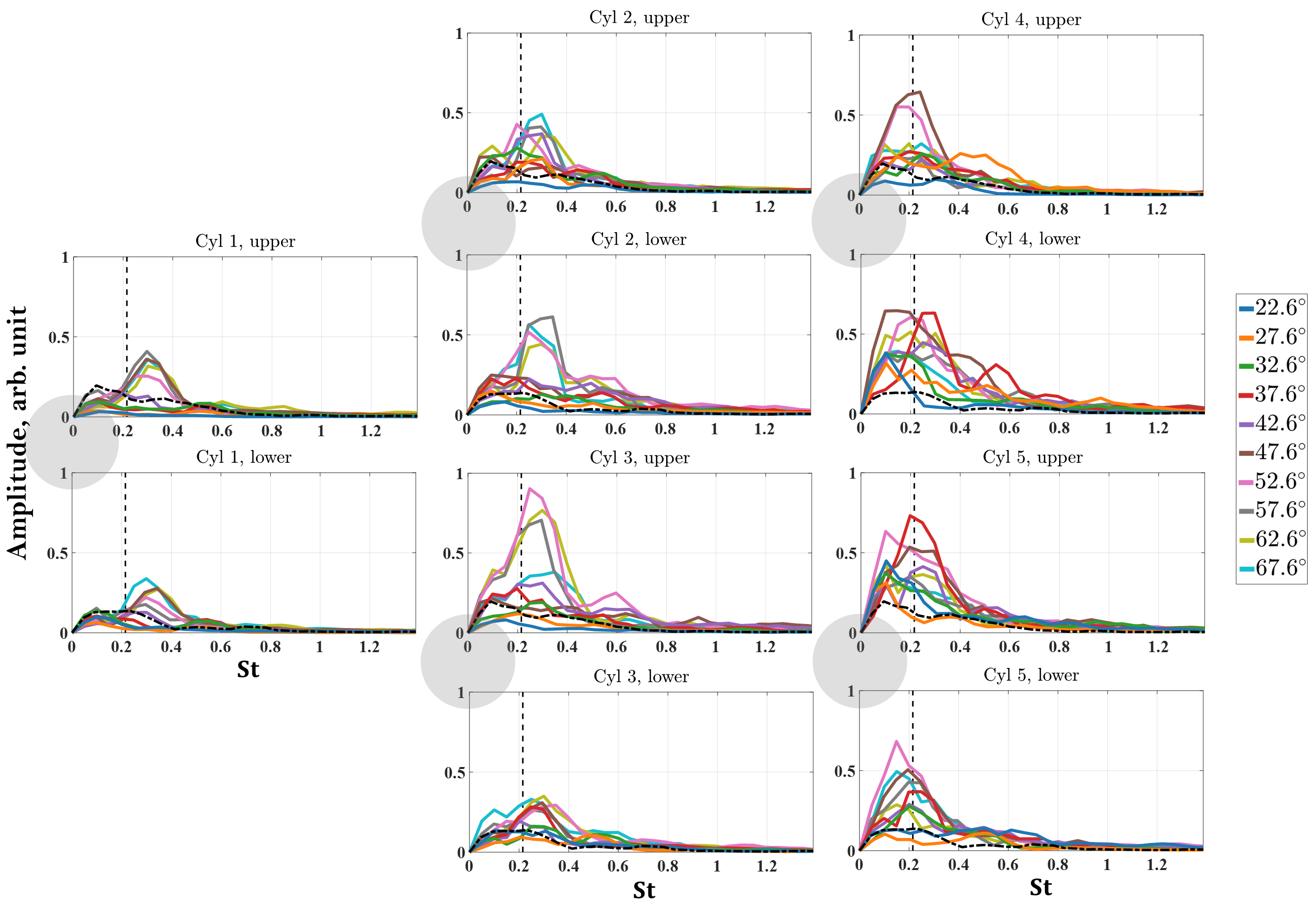}
    \caption{Frequency spectra of the vortex shedding based on circulation, $\Gamma$, in a $d \times d$ square region $1.25d$ downstream of the upper and lower shoulders (center of the square regions are downstream of the uppermost and lowermost points of the cylinder) of each cylinder member for all the $\pmb \vee$-formations. The dashed vertical black line indicates the shedding frequency for an ideal long circular cylinder. For visualization purpose, we use light gray circles to qualitatively denote the relative position of the formation members and connect the two plots of the frequency spectra of the upper and lower circulations for the same member. The black dash-dotted lines correspond to the frequency spectra of the vortex shedding (based on circulation) behind a single, isolated cylinder.}\label{fig:shedspec2}
\end{figure}

From Fig.~\ref{fig:circ}, we see that the leading and interior members for the first two most overlapping formations ($\phi = 22.6^{\circ}$ and $27.6^{\circ}$) show an almost flat non-periodic variation of $\Gamma$ in their wakes. This is due to the trapping of recirculation regions and wakes of the leading and interior members in the separation bubbles  for these tight formations as discussed in Sec.~\ref{sec:global} and Sec.~\ref{sec:local} and visible in Fig.~\ref{fig:streamlines}(a-b). For the $\phi = 22.6^{\circ}$ case, the $\Gamma$ variations in the wakes of members 2 and 3 on the interior side (`Cyl 2 lower' and `Cyl 3 upper' in Fig.~\ref{fig:circ}) are identical, due to the coinciding/overlapping closed curves used to compute $\Gamma$. As $\phi$ increases, the wake behind the members have more room to develop further and shed, and we see larger variations in circulation owing to periodic vortex shedding. Especially for the non-overlapping formations, we see larger amplitudes of oscillation in the circulation calculated on the interior side of the formation (lower side of cylinders 2 and 4, and upper side of cylinders 3 and 5) compared with those outside the formation. Among the intermediate overlapping cases, formation with $\phi = 32.6^{\circ}$ also shows lower or no oscillation similar to that of the most overlapping cases, while from $\phi = 37.6^{\circ}$ forward we start to see the amplitudes of oscillation growing with similar patterns as that of the non-overlapping cases. Starting from $\phi = 37.6^{\circ}$ and higher, we see the mean of the $\Gamma_{\rm upper}$ and $\Gamma_{\rm lower}$ for each member to be getting close to each other and hovering around the circulation values of the single cylinder case, leading to minimal or no lift on the bodies. This is especially the case for the trailing members of all formations with $\phi \geqslant 37.6^{\circ}$.

As mentioned above, the sum of the circulations from both the upper and lower shoulders can give an indication of the lift force experienced by a member. For interior and trailing members (2, 3, 4, and 5) of formations with $\phi \leqslant 32.6^{\circ}$, the largest variations are seen in the $\Gamma$ calculated in the interior of the formation (lower for cylinders 2/4 and upper for cylinders 3/5). In these formations the inner shear layers of the members are mostly controlled by the bleeding flow while the outer shear layers are free to move and expand as needed (as discussed in Sec.~\ref{sec:global} and \ref{sec:local}). Therefore, on averaging and taking a sum, the circulation magnitude (i.e. absolute value) on the outer shoulder is greater in general. For example, looking at cylinder 2, the average of $\Gamma / (U_{\infty} d)$ from the lower (inner) shoulder for all formations with $\phi \leqslant 37.6^{\circ}$ is less than 1 whereas that for the upper (outer) shoulder is around 1.5 (for all these four formations). This indicates the  lift force experienced by cylinder 2 in these formations is in $+y$ direction (outward). Similarly, most members of these 4 formations with the largest recorded flow asymmetries (except the leading members) experience a small lift force (lateral force) in the outward direction (away from $y=0$). But as discussed previously, these are small compared to the streamwise drag forces (along $+x$ direction). Note that the variations in the phase of the plots are related to variations of the timing of consecutive-overlapping experimental data collections and the results have not been time-matched.

In addition to the average magnitude of the circulation that is affected by the presence of multiple members, as seen in Fig.~\ref{fig:circ}, the frequency of the vortex shedding is also impacted. To further explore the temporal characteristics of vortex shedding, we present the frequency spectra of vortex shedding in Fig.~\ref{fig:shedspec2}, by taking the fast Fourier transform (FFT) of the time-series of circulation presented in Fig.~\ref{fig:circ}. Circulation, with its integral definition serves as an average behavior of a region and is chosen over using a single point or few/many points in the wake. The frequency, $f$, is presented in the form of dimensionless Strouhal number (${\rm St} = fd/U_{\infty}$). The dashed vertical black line indicates the shedding frequency for an ideal long circular cylinder. 

For formations with small $\pmb \vee$-angles (most overlapping cases), significant interactions between the array members lead to irregular vortex shedding, leading to a more broad-band frequency response where for members 1, 2, and 3, it is difficult to identify a main peak frequency. However, for the trailing members 4 and 5, on the interior of the formation, we record a peak frequency in the similar range as that of the single cylinder case. The more pronounced frequency responses are recorded for non-overlapping cases where we see an increase in the shedding frequency on both interior and exterior sides of leading member 1, and interior members 2 and 3. The trailing members, however, experience lower shedding frequencies than the the rest of the members, but the frequencies are still larger than that of the single cylinder case. The formations with intermediate overlaps show frequencies in between the two other groups, with combinations of both broadband responses and peak frequency increases depending on the asymmetry of the flow around each of the bodies. 

While the main focus of this work has been on the impact of formations on drag force, the discussion presented here along with Figs.~\ref{fig:circ} and ~\ref{fig:shedspec2} show the potential of formations to be used for tuning multiple flow characteristics in addition to the drag. Future works should also focus on the exploration of the mechanisms of the variations in the temporal flow response past each member of the formation as well as the group as whole, and how the variations in the geometry of the formation can impact the frequency response and the lift force experienced by the bodies and consider that as part of the optimization process to be used for formations of uncrewed vehicles. 

\section{Conclusions}\label{sec:conclusion}

As demonstrated here, $\pmb{\vee}$ formations are an effective way to control the flow field to tune the drag force of each member of a group and the entire formation. Over a wide range of formation angles, we see various levels of drag reduction, from substantial reductions as much as 78\% (member 3 of the $\phi = 22.6^{\circ}$) in formations with overlaps in frontal views, to moderate ones for formations with intermediate or no-overlaps. At the streamwise spacing chosen (slightly less than formation length of a single cylinder), we see that up to an angle of around $50^{\circ}$ we can achieve variations of drag reduction depending on the position of the member. Beyond this angle, we only see the leading member experiencing a drag reduction. 

The interactions of the flow between the members can be approached as a combination of wake-body and wake-wake interactions and how the bleeding flow (or gap flow) regulates these interactions. By separating the streamlines into groups according to their trajectories, we further explore how the behavior of the shear layer attachment/detachment and the wake interactions modulate the pressure and friction drags of each member qualitatively which result in the quantitative drag values presented. We also show variations in the turbulent kinetic energy within the wakes of all the bodies, compared with a solo-cylinder, where formations with overlaps limit the transition to turbulence, while the non-overlapping formations reach similar levels of turbulent kinetic energy as that of the wake of a single cylinder. 

The regulating nature of formations and their geometry is clearly visible in the variations in the velocity fields and pressure distributions between the members, as shown in the plots of velocity and pressure profiles upstream and downstream of each member in Sec.~\ref{sec:local}. In formations with overlaps, members can experience significant levels of pressure recovery which has a direct impact on the drag force experienced by the bodies. We also see that grouping the formations based on their level of overlap has allowed us to describe the mechanism of the flow behavior in a more systematic manner, however, as shown the borderline between the groups are not fully fixed and depending on the phenomenon of interest (drag, shedding, pressure field, etc.) the formation angles included in groups need to vary. 

Additionally, the pressure variations and the asymmetry between the top and bottom sides of each of the cylinders, especially leading and interior members (1, 2, and 3) of these formations with overlaps can also result in changes in the lift experienced by the body where varying levels of circulations are recorded in the wakes of each of the bodies. We also see potential in variations in the vortex shedding frequencies in the wakes of each of the members, especially for formation with no-overlaps. 

The results of this study provide a baseline for further investigation of the wake-body and wake-wake interactions of flow past groups of bodies. It is clear that the choice of the separation and orientation between the bodies has a direct impact on the aero-/hydrodynamic performance in terms of drag; however, as briefly discussed, there is further room for expanding the investigations to the temporal flow behavior and the potential for altering the lift forces of the bodies and the vortex shedding frequency, as well as considerations for the stability of the bodies, vortex induced vibration, and the impact on the moments experienced by non-symmetric bodies in groups. In addition, the similarities observed between the asymmetric wake fields in tight formations with smaller angles, and the fluidic pinball experiments and flow past side-by-side cylinders merit further experimental and theoretical investigations to study the stability of the flow past formations with tight $\pmb \vee$ angles and the impact of the stability on the aero-/hydro-dynamic loading on the bodies and will be a future work. While this work only gives insight on potential paths one can explore for tuning the drag of each member and the entire formation, ultimately, a combined understanding of all the flow features impacted by the formation geometry, including drag, lift, and the temporal flow behavior will be required to inform the formation designs needed to meet the objectives of the aerial/under-water vehicles of interest along with the necessary considerations for safety and collision-avoidance of aerial/under-water vehicles in tight formations.   

\bibliography{library}

\newpage

\pagenumbering{arabic}
\makeatletter 
\renewcommand{\thefigure}{SM.\@arabic\c@figure}
\renewcommand{\thetable}{SM.\@arabic\c@table}
\makeatother
\setcounter{figure}{0}

\section*{Supplemental Material}

{\begin{center} \Large \bf Impact of the formation angle on the drag of bio-inspired $\pmb \vee$-formations
\end{center}}
{\begin{center} by Prasoon Suchandra, and Shabnam Raayai-Ardakani\end{center}}

\begin{figure}[!ht]
    \centering
    \includegraphics[width=1\linewidth]{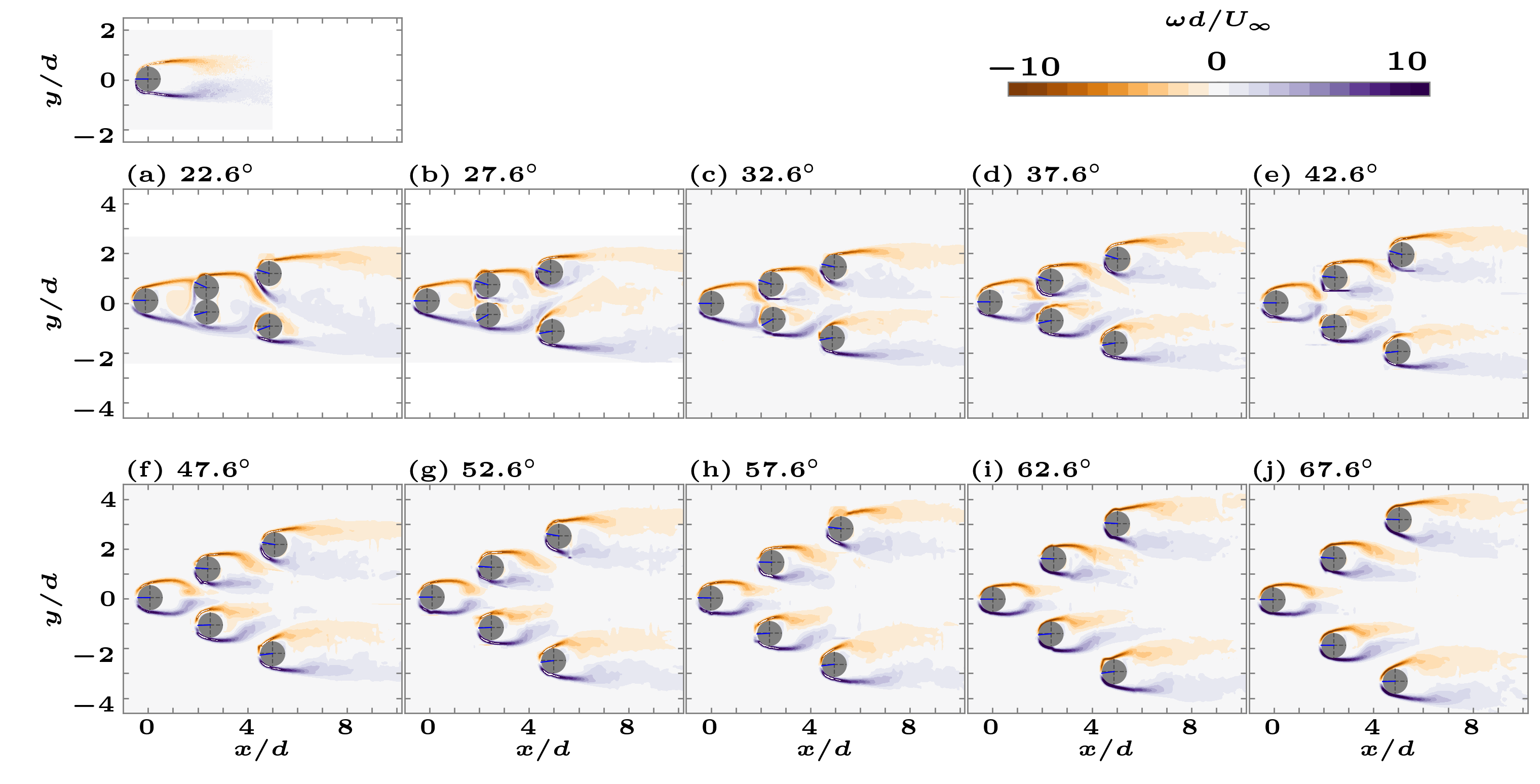}
    \caption{Contours of normalized mean vorticity, $\omega(x,y)d/U_{\infty}$ for formations with $\phi$ going from $22.6^{\circ}$ to $67.6^{\circ}$ in steps of $5^{\circ}$. Sub-figures (a-e) correspond to the cases with frontal overlaps and sub-figures (f-j) correspond to the cases with no frontal overlaps. The vorticity contours for the solo cylinder is presented on the top as reference.}
    \label{fig:Omega}
\end{figure}

\begin{table}[!ht]
\begin{center}
\begin{tabular}{c|c|c|c|c|c|c}
 \ \ \  $\phi$ \ \ \ & \ \ \ $U_{12}^b/U_{\infty}$ \ \ \ & \ \ \ $U_{13}^b/U_{\infty}$ \ \ \ & \ \ \ $U_{23}^b/U_{\infty}$ \ \ \ & \ \ \ $U_{24}^b/U_{\infty}$ \ \ \ & \ \ \ $U_{35}^b/U_{\infty}$ \ \ \ & \ \ \ $U_{45}^b/U_{\infty}$ \ \ \ \\ \hline
$22.6^{\circ}$ & 0.029 & (-)0.027 & 0.000 & 0.193 & 0.147 & 0.498 \\ 
$27.6^{\circ}$ & 0.088 & (-)0.003 & 0.289 & 0.185 & 0.263 & 0.542 \\ 
$32.6^{\circ}$ & 0.109 & 0.060 & 0.385 & 0.221 & 0.266 & 0.599 \\ 
$37.6^{\circ}$ & 0.174 & 0.149 & 0.676 & 0.324 & 0.356 & 0.684 \\ 
$42.6^{\circ}$ & 0.212 & 0.221 & 0.629 & 0.313 & 0.348 & 0.672 \\ 
\hline
$47.6^{\circ}$ & 0.354 & 0.311 & 0.928 & 0.375 & 0.467 & 0.775 \\ 
$52.6^{\circ}$ & 0.388 & 0.381 & 0.943 & 0.480 & 0.539 & 0.835 \\ 
$57.6^{\circ}$ & 0.487 & 0.472 & 0.991 & 0.478 & 0.496 & 0.802 \\ 
$62.6^{\circ}$ & 0.568 & 0.475 & 0.931 & 0.553 & 0.560 & 0.833 \\ 
$67.6^{\circ}$ & 0.602 & 0.650 & 1.041 & 0.592 & 0.564 & 0.853 \\ 
\end{tabular}
\caption{Bleeding flow magnitudes between the members of different $\pmb \vee$-formations. Overlapping and non-overlapping formations are separated by a line in the middle of the table. (-) sign indicates the bleeding flow is from inside to outside the array.}  \label{tab:bleed}
\end{center}
\end{table}

\end{document}